\documentclass[10pt]{article}

\usepackage{amsmath}
\usepackage{amssymb}  
\usepackage[T1]{fontenc}
\usepackage{pifont}
\usepackage{pstricks}
\usepackage{amsfonts}
\usepackage{graphicx}
\usepackage{amsmath}
\usepackage{authblk}
\usepackage{pstricks} 
\usepackage{pstricks-add}
\usepackage{caption}
\usepackage{placeins}
\usepackage{pstricks}
\usepackage{pdfpages}
\usepackage{framed}
\usepackage{cancel}
\usepackage{bm}
\usepackage{epstopdf}
\def\x2dot{\mathop{x}\limits}
\def\y2dot{\mathop{y}\limits}

\def\bfy2dot{\mathop{\bf y}\limits}
\def\z2dot{\mathop{z}\limits}

\def\csi2dot{\mathop{\xi}\limits}
\def\et2dot{\mathop{\eta}\limits}
\def\bet2dot{\mathop{\beta}\limits}
\def\t2dot{\mathop{\theta}\limits}
\def\s2dot{\mathop{\sigma}\limits}
\def\d2dot{\mathop{\delta}\limits}
\def\q2dot{\mathop{q}\limits}
\def\l2dot{\mathop{\lambda}\limits}
\def\ps2dot{\mathop{{\cal E}}\limits}
\def\tet2dot{\mathop{\theta}\limits}
\def\bfx2dot{\mathop{\bf x}\limits}
\def\bfy2dot{\mathop{\bf y}\limits}
\def\bfq2dot{\mathop{\bf q}\limits}
\def\bbfq2dot{\mathop{\bar {\bf q}}\limits}
\def\w2{\mathop{W}\limits}
\def\xgrande2dot{\mathop{\bf X}\limits}
\def\p02dot{\mathop{P}\limits}
\def\a2dot{\mathop{A}\limits}
\def\vett{\mathop{\bf x}\limits}

\newtheorem{prop}{Proposition}

\newtheorem{rem}{Remark}
\newtheorem{exe}{Example}

\title{A simple approach to nonlinear nonholonomic systems with several examples}
\author{F.~Talamucci}
\affil{{\it DIMAI, Dipartimento di Matematica e Informatica ``Ulisse Dini''},\\
{\it	Universit\`a degli Studi di Firenze, Italy}\\
{\it	e-mail: federico.talamucci@unifi.it}}
\date{}

\begin{document}
	\bibliographystyle{plain}
	
	\setcounter{equation}{0}

	\maketitle
	
	\vspace{.5truecm}
	
	
	\vspace{.5truecm}
	
	
	\vspace{.5truecm}
	

\begin{abstract}
	
	\noindent
	The main theme of the article is the study of discrete systems of material points subjected to constraints not only of a geometric type (holonomic constraints) but also of a kinematic type (nonholonomic constraints).
	The setting up of the equations of motion follows a simple principle which generalizes the holonomic case. Furthermore, attention is paid to the fact that the kinematic variables retain their velocity meaning, without resorting to the pseudo-velocity technique.
	Particular situations are examined in which the modeling of the constraints can be carried out in several ways to evaluate their effective equivalence.
	Numerous examples, many of which taken from the most recurring ones in the literature, are provided in order to illustrate the proposed theory.
\end{abstract}

\section{Introduction}

\noindent
The definition of holonomic or geometric constraint is well known as a position constraint, 
i.~e.~analytically characterized by the annulment of a function of the coordinates alone.
In turn, a system subject to holonomic constraints is classified as scleronomous if the constraints do not explicitly depend on time or rheonomous if they do not.
The definition of a nonholonomic system is more complex, if in this term we want to gather all the complementary situations with respect to the case of geometric restrictions: in fact the term nonholonomic refers to all those restrictions which cannot be expressed solely through a relationship between the spatial coordinates that fix the positions of the system.
In the broad scenario of restrictions of this type we focus our attention on the constraints expressible through relations involving the coordinates of the points and their velocities: this category of nonholonomic constraints certainly concerns a wide field of applications in mechanical and engineering problems (for example in robotics), beyond than to be relevant from a point of view of the theory of motion.

\noindent
The main purpose of the work is to provide a simple method to address the formulation of the mathematical model of a nonholonomic system, mainly having in mind two aspects:
\begin{itemize}
\item[$(a)$] 
to extend as much as possible the fundamental features of the known and consolidated theory of holonomic systems,
\item[$(b)$]
to mantain the real velocities as the kinetic variables, placing in second order the need to adopt a set of pseudovariables for the analytical resolution of the problem.
\end{itemize}

\noindent
As for the second aspect, there is no doubt that the introduction within a specific problem of suitable combinations of velocities (these variables may no longer represent velocities and are therefore called pseudovelocities) drastically simplifies the mathematical solution of the set of equations; notable examples in this sense are the pure rolling of disks or spheres on the plane or a surface (we quote \cite{bloch1}, \cite{bloch2} and the cited literature).
On the other hand, there is a considerable series of problems associated with kinematic constraints whose natural formulation occurs through Cartesian coordinates or in any case through variables which correspond directly to the velocities, without any transformation (we denote this situation by true coordinates systems).
This setting also makes the procedure for deducing energy-type information from the motion equations more natural.

\noindent
As far as point $(a)$ is concerned, a very brief review is needed to frame the historical development of the study of nonholonomic systems. 

\noindent
It is well known that the theory of holonomic systems has its foundation in the analytic fromalism of the end of the eighteenth century by Euler and Lagrange. 
In the following years a series of simple examples (such as the pure rolling of a rigid body on a plane) brought attention to the fact that, if it is true that a geometrical restriction entails a precise restriction on the speed of the system, the converse it is not necessarily true: a constraint on the possible speeds does not necessarily imply a restriction on the possible configurations.
This has led to consider a new category of constraints which consists in formulating a relationship between the coordinates and the velocity variables. If this relation cannot be reported in terms of the coordinates only (in this case the constraint is said to be integrable), then we are in the presence of an nonholonomic constraint.

\noindent
The various types of equations proposed for nonholonomic systems initially concern linear kinematic constraints with respect to velocities:
in fact, examples and applications concern only linear nonholonomic constraints and the main and historical proposals of formal equations, edited by  ${\rm {\check C}}$aplygin, Hamel, Maggi, Appell, Voronec
and others, are masterfully reported and commented on in the fundamental work by Neimark and Fufaev on this subject \cite{neimark}.

\noindent
From the geometric point of view, the condition of a nonholonomic system with linear constraints is strongly analogous to that of a holonomic system:
the space of the configurations, established by the Lagrangian coordinates, remains the same, while the space of the admitted velocities, instead of being the entire vector space of dimension equal to the degrees of freedom, will be a linear subspace of it.
In this way, writing Newton's law along the directions admitted by the constraints leads to equations of the Lagrangian type which are the obvious extension of the holonomic case.

\noindent
The situation regarding nonlinear kinematic constraints is decidedly different:
we can trace the first example of realization of a mechanical system with nonlinear kinematic constraints in the publication \cite{appell2} and more recently analyzed, among others, by \cite{hamel}, \cite{neimark} and \cite{zek1}.
Among the elements that animate the debate on the implementation of nonlinear constraints there is the frequent situation of multiple possibilities to give rise to the same constraint constraints with nonlinear expressions or with a set of equivalent linear expressions.
It must also be said that from a theoretical point of view there are not many texts on analytical mechanics that deal with the theoretical formulation of nonlinear nonholonomic systems: even the most notable texts on nonholonomic systems (\cite{lurie}, \cite{pars}) provide an exhaustive theory for the linear case, formulated via pseudo coordinates.
A notable exception is the text \cite{papastrav}, which offers a valuable overview of the study of nonlinear nonholonomic systems and formulates the theory of motion for them, using several methods.

\noindent
In general terms we can identify at least three distinct methods for formulating the equations of motion
(even if they can evidently interact):
\begin{itemize}
\item [$(i)$] a procedure based on the analysis of the displacements admitted to the system by the constraints and on the generalization of the d'Alembert's principle,
\item [$(ii)$] methods based on a variational principle or on the use of multipliers,
\item [$(iii)$] an approach based on the formalism of differential geometry.
\end{itemize}

\noindent
As for the latter, we can trace the geometric treatment of constrained mechanical systems in \cite{vagner}, \cite{vranceanu} the pioneering works and indicate, among others, in \cite{leon}, \cite{krup}, \cite{saleh} the formal complexity of this theoretical sector which has promoted considerable progress in differential geometry.
An excellent publication that strikes a balance between the formal presentation of theory and the development of practical examples and problem solving is \cite{swac}. 
Regarding $(ii)$, a significant reference that contains, among other things, the main bibliography on the subject is \cite{ranada}.
As for the question of making the equations of motion of a nonlinear nonholonomic system derive from a variational principle, the problem is still open under various aspects; an extensive and timely discussion of these issues is in a significant reference that contains the main bibliography about it is
\cite{flannery}.

\noindent
In the present work we turn to a method of type $(i)$,
basing ourselves on the possible displacements compatible with the constrained restrictions and generalizing the well-known equation formulation procedure through the so-called virtual work principle.
The structure of the work is very simple: in Section 2 we deal with the geometric and kinematic aspects of constrained systems, admitting a very generic class of constraints. Various examples proposed develop the theory presented.
In Section 3 the equations of motion for nonholonomic systems with nonlinear constraints are formulated and various comments and observations are added regarding particular cases or specific hypotheses.
Also in this part examples of systems with associated equations of motion are proposed.

\section{The mathematical model}
 
\subsection{A general class of constraints}

\noindent
On the basis of the selected approach method to formulate the motion and the typology of examples that we have in mind, it is better to use the Lagrangian starting setting of a discrete material system formed by a finite number of material points.
Let us therefore  consider any system of $N$ particles of mass $m_i$ which are located in an inertial frame of reference by the coordinates ${\bf x}_i=(x_i, y_i, z_i)$, $i=1, \dots, N$.
The Newton's equation 
$$
m_i \vett^{..}= {\bf F}_i+{\bm \Phi}_i \qquad i=1, \dots, N
$$ 
where ${\bf F}_i$ and ${\bm \Phi}_i$ are respectively the active and the constraint forces on the $i$-th particle, can be summarized in ${\Bbb R}^{3N}$ by writing 
\begin{equation}
\label{newton}
{\dot {\bf Q}}={\bf F}+{\bm \Phi}
\end{equation}
with ${\bf Q}=(m_1 {\dot {\bf x}}_1, \dots, m_N{\dot {\bf x}}_N)\in \Bbb R^{3N}$ is the vector listing the linear momenta,  
${\bf F}=({\bf F}_1, \dots, {\bf F}_N)\in {\Bbb R}^{3N}$, 
${\bm \Phi}=({\bm \Phi}_1, \dots, {\bm \Phi}_N)\in {\Bbb R}^{3N}$ the vectors representing all the active forces and the constraint forces, respectively.

\noindent
The unknown forces ${\bm \Phi}$ in (\ref{newton}) are due to restrictions enforced to the configurations and to the kinematics the system: we will assume here that the constraints can be formulated by the $r<3N$  equations:
\begin{equation}
\label{eqvinc}
\left\{
\begin{array}{l}
f_1({\bf X}, {\dot {\bf X}}, t)=0, \\
\quad \dots \quad \dots \quad \dots \\
\quad \dots \quad \dots \quad \dots \\
f_r({\bf X}, {\dot {\bf X}},t)=0
\end{array}
\right.
\end{equation}
where for brevity we set ${\bf X}=({\bf x_1}, \dots, {\bf x}_N)\in {\Bbb R}^{3N}$, ${\dot {\bf X}}=({\dot {\bf x}}_1, \dots, {\dot {\bf x}}_N)\in {\Bbb R}^{3N}$.
Hence, the constraints we will consider here concern conditions that rescrict the positions (through ${\bf X}$) and the velocities (through ${\dot {\bf X}}$) of the particles and they possibly depend explicitly on time $t$ (moving or rheonomic constraints). When $t$ is absent from (\ref{eqvinc}), the constraints are said fixed or scleronomic.

\noindent
The conditions (\ref{eqvinc}) are assumed to be independent with respect to the kinematical variables ${\dot {\bf X}}$, meaning that the $r$ vectors in ${\Bbb R}^{3N}$
$$
\nabla_{\dot {\bf X}}f_i=\left(\dfrac{\partial f_i}{\partial {\dot x}_1}, \dots, 
\dfrac{\partial f_i}{\partial {\dot z}_N}\right), \qquad i=1, \dots, r
$$
are linearly independent or, equivalently, the jacobian matrix formed by the $r$ vectors 
has the full rank:
\begin{equation}
	\label{rankr}
	rank\;J_{\dot {\bf X}}(f_1, \dots, f_r)=r
\end{equation}

\noindent
We don't rule out the possiblity that one or more of the constraints are actually geometrical conditions: this occurs if $f_j$ is linear w.~r.~t.~${\dot {\bf X}}$ and it exists a function $\psi_j({\bf X},t)$ such that 
\begin{equation}
\label{fjint}
\dfrac{d\psi_j}{dt}=\nabla_{\bf X} \psi_j\cdot {\dot {\bf X}}
+\dfrac{\partial \psi_j}{\partial t}=f_j({\bf X}, {\dot {\bf X}},t)
\end{equation}

\noindent
We settle such an occurrence by assuming that the first $h$ conditions in (\ref{eqvinc}) are indeed integer conditions (holonomic constraints),
whereas the last $h$ are purely kinematic (nonintegrable) constraints and entail the nonholonomic state of the system. 

\noindent
The validity of (\ref{fjint}) for $j=1, \dots, h$ and the regularity assumption (\ref{rankr}), which in turn implicates that the vectors $\nabla_{\bf X}\psi_j=\nabla_{\dot {\bf X}}f_j$, $j=1, \dots, h$ are linearly independent, allow us to acquire the $n=3N-h$ lagrangian coordinates $q_1$, $\dots$, $q_n$ 
from the system $\psi_1({\bf X},t)=0$, $\dots$, $\psi_h({\bf X},t)=0$ by achieving the $3N$ relations
$x_1=x_1(q_1, \dots, q_n, t)$, $\dots$, $z_n=z_n(q_1, \dots, q_n, t)$ summarized by 
\begin{equation}
\label{xqt}
{\bf X}={\bf X}({\bf q}, t), \qquad {\bf q}=(q_1, \dots, q_n)
\end{equation}

\noindent
The calculation of the derivative of (\ref{xqt}) provides the following representation of the velocity of the system
\begin{equation}
	\label{vellagr}
	{\dot {\bf X}}=\sum\limits_{i=1}^n \dfrac{\partial {\bf X}}{\partial q_i}{\dot q}_i+\dfrac{\partial {\bf X}}{\partial t}={\dot {\bf X}}({\bf q}, {\dot {\bf q}},t)
\end{equation}
which is pertinent to holonomic systems: in the present case, the additional kinematic conditions 
$f_{h+1}=0$, $\dots$, $f_r=0$ in (\ref{eqvinc}) (which we assumed to be not integrable) have to be taken into consideration. 
In light of this, it is suitable to define, by means of (\ref{vellagr}),
\begin{equation}
\label{fij}
\phi_j({\bf q}, {\dot {\bf q}},t)=f_{h+j}({\bf X}({\bf q},t), {\dot {\bf X}}({\bf q}, {\dot {\bf q}},t),t)\qquad \textrm{for each}\;\;\;j=1,\dots, k=r-h
\end{equation}
and to rewrite the last $k$ conditions in (\ref{eqvinc}) as
(\ref{vellagr}),

\begin{equation}
	\label{constr}
	\begin{cases}
		\phi_1(q_1,\dots, q_n, {\dot q}_1, \dots, {\dot q}_n, t)=0\\
		\qquad \qquad \dots  \dots  \dots\\
		\qquad \qquad \dots \dots \dots\\
		\phi_k(q_1,\dots, q_n, {\dot q}_1, \dots, {\dot q}_n, t)=0
	\end{cases}
\end{equation}

\begin{rem}
Although the setting in the environment space ${\Bbb R}^{3N}$ facilitates the determination of the compatible velocities in (\ref{dalamb}), the starting point of the problem might well be 
any mechanical system identified by the $n$ Lagrangian free coordinates $q_1, \dots, q_n$ 
and subject to the kinematical conditions (\ref{constr}): the same conclusions we will present in the following analysis apply also in this general case.
\end{rem}

\noindent
Regarding (\ref{constr}), it is worth to underline that the constraints are independent without any extra assummption:
\begin{equation}
	\label{vincind}
	rank\,J_{({\dot q}_1, \dots, {\dot q}_n)}(\phi_1, \dots, \phi_k)=k.
\end{equation}
Actually, one has owing to (\ref{vellagr}) and (\ref{fij}):
$$
J_{	\dot {\bf q}}(\phi_1, \dots, \phi_k)=
J_{\dot {\bf X}}(f_{h+1}, \dots, f_r)\,J_{\dot {\bf q}}{\dot {\bf X}}=
J_{\dot {\bf X}}(f_{h+1}, \dots, f_r)\,J_{\bf q}{\bf X}
$$
The first jacobian matrix has the full rank $k$ (because of (\ref{rankr})) and the second the rank $n$ (because of the indipendence of the vectors $\frac{\partial {\bf X}}{\partial q_i}$, $i=1, \dots, n$), so that (\ref{vincind}) is fulfilled.

\noindent
The latter condition does so that (\ref{constr}) can be locally written explicitly with respect to a selection of $n-k$ variables: with no loss in generality we can assume that the nondegenerate square matrix of order $k$ is formed by the last $k$ columns, so that 
$det\,J_{({\dot q}_{n-k+1}, \dots, {\dot q}_n)}(\Phi_1, \dots, \Phi_k)\not =0$ and the explicit equations deduced from (\ref{constr}) are
\begin{equation}
	\label{constrexpl}
	\begin{cases}
		{\dot q}_{m+1}=\alpha_1(q_1, \dots, q_n, {\dot q}_1, \dots, {\dot q}_m, t) \\
		\dots  \\
		\dots  \\
		{\dot q}_n=\alpha_k(q_1, \dots, q_n, {\dot q}_1, \dots, {\dot q}_m, t)
	\end{cases}
\end{equation}
with $m=n-k$. The parameters ${\dot q}_1, \dots, {\dot q}_m$ are now playing the role of the basic and the independent velocities; in any position $(q_1, \dots, q_n)$, the selection of a $m$--uple of such parameters in ${\Bbb R}^m$ defines completely the kinematic state of the system: indeed, by virtue of (\ref{constrexpl}) the velocity (\ref{vellagr}) takes the form
\begin{equation}
	\label{vellagrm}
	{\dot {\bf X}}=\sum\limits_{r=1}^m \dfrac{\partial {\bf X}}{\partial q_r}{\dot q}_r+
	\sum\limits_{\nu=1}^k \dfrac{\partial {\bf X}}{\partial q_{m+\nu}}\alpha_\nu+
	\dfrac{\partial {\bf X}}{\partial t}={\dot {\bf X}}(q_1, \dots, q_n, {\dot q}_1, \dots, {\dot q}_m,t)
\end{equation}
Essentially, each of the $h$ integer constraints in (\ref{eqvinc}) subctracts a degree of freedom from the $3N$ coordinates ${\bf X}$ leading to $n=3N-h$ independent coordinates $q_1$, $\dots$, $q_n$; on the other hand, each of the $k$ kinematic constraints removes one of the velocities ${\dot q}_1$, $\dots$, ${\dot q}_n$, so that only $m=n-k$ have to be considered independent. 

\subsection{Some examples of nonlinear nonholonomic systems}

\noindent
We find it convenient to introduce some concrete models, selecting from the most recurrent ones in literature. 
		\begin{exe}
		A particle $P$ moves in the space ${\Bbb R}^3$ in respect of the following condition on the velocity:
		\begin{equation}
			\label{pct}
			|{\dot P}|=C(t)
		\end{equation}
		with $C$ nonnegative function of time. The case $C$ constant is frequently debated in literature (ammong others \cite{virga}, \cite{swac}); the case $C(t)=1/\sqrt{t}$ is analyzed in \cite{swac}, \cite{krup}. In a coordinate system (\ref{pct}) writes
		${\dot x}^2+{\dot y}^2+{\dot z}^2-C(t)=0$ and according to (\ref{eqvinc}) it is $h=0$ (no geometric restriction), $k=r=1$ and (\ref{xqt}) is simply $(x,y,z)=(q_1, q_2, q_3)$, so that
		in (\ref{constr}) $n=3$ and (\ref{constrexpl}) is ($m=1$) 
		$$
		{\dot q}_3=\pm \sqrt{C^2(t)-({\dot q}_1^2+{\dot q}_2^2)}.
		$$
		More generally, in (\ref{pct}) can be considered part of the following kinematic condition, involving each singular component:
		$$
		a{\dot x}^2+b{\dot y}^2+c{\dot z}^2=C^2(t)
		$$
		The case $a=b$, $c=-1$, $C(t)=0$ is examined in \cite{fasso}. 
	\end{exe}
	
	\begin{exe}
		Let us consider $N$ points moving in the space without geometric constraints while keeping 
		the same magnitude of the velocity:
		$$
		|{\dot P}_1|=|{\dot P}_2|=\dots=|{\dot P}_N|
		$$
The case $N=2$ is studied in \cite{virga}; for a general integer $N\geq 2$ equations (\ref{eqvinc}) correspond to the $N-1$ kinematic conditions 
		
	\begin{equation}
		\label{samevel}
		\left\{
		\begin{array}{l}
			{\dot x}_1^2+{\dot y}_1^2+{\dot z}_1^2-\left({\dot x}_2^2+{\dot y}_2^2+{\dot z}_2^2\right)=0, \\
			\\
		{\dot x}_1^2+{\dot y}_1^2+{\dot z}_1^2-\left({\dot x}_3^2+{\dot y}_3^2+{\dot z}_3^2\right)=0, \\
		\\
		\dots \quad \dots \quad \dots \quad \dots \\
			\dots \quad \dots \quad \dots \quad \dots \\
			\\
		{\dot x}_1^2+{\dot y}_1^2+{\dot z}_1^2-\left({\dot x}_N^2+{\dot y}_N^2+{\dot z}_N^2\right)=0,
	\end{array}
\right.
\end{equation}
which are evidently independent (in the sense of (\ref{rankr})) whenever the velocity in common is not null.
In this problem $k=r=N-1$ and $n=3N$, thus $m=n-k=2N+1$, which is the number of independent kinetic variables (selected among ${\dot x}_1$, ${\dot y}_1, \dots, {\dot z}_N$) required to express the remaining $N-1$ variables; a reasonable selection of the $2N+1$ independent velocities is
$$
(\overbrace{{\dot q}_1, {\dot q}_2, {\dot q}_3}^{3},
\overbrace{{\dot q}_4, {\dot q}_5, {\dot q}_6, {\dot q}_7, \dots, 
	{\dot q}_{2N}, {\dot q}_{2N+1}}^{2(N-1)})=(
\overbrace{{\dot x}_1, {\dot y}_1, {\dot z}_1}^{3}, \overbrace{{\dot x}_2, {\dot y}_2, {\dot x}_3, {\dot y}_3, \dots, {\dot x}_N, {\dot y}_N}^{2(N-1)})
$$
and the $N-1$ explicit equations (\ref{constrexpl}) for the dependent velocities 
$({\dot q}_{2N+2}, {\dot q}_{2N+3}, \dots, {\dot q}_{3N})=({\dot z}_2, {\dot z}_3, \dots, {\dot z}_N)$ are
\begin{equation}
	\label{samevelexpl}
	\left\{
	\begin{array}{l}
{\dot z}_2=	{\dot q}_{2N+2}=\alpha_1=\pm\sqrt{{\dot q}_1^2+{\dot q}_2^2+{\dot q}_3^2
			-{\dot q}_4^2-{\dot q}_5^2} \\
		\\
{\dot z}_3={\dot q}_{2N+3}=\alpha_2=\pm\sqrt{{\dot q}_1^2+{\dot q}_2^2+{\dot q}_3^2-{\dot q}_6^2-{\dot y}_7^2} \\
		\\
		\quad \dots \quad \dots \quad \dots \\
		\\
{\dot z}_N={\dot q}_{3N}=\alpha_{N-1}=\pm\sqrt{{\dot q}_1^2+{\dot q}_2^2+{\dot q}_3^2-{\dot q}_{2N}^2-{\dot y}_{2N+1}^2}
	\end{array}
	\right.
\end{equation}
\end{exe}
			
	\begin{figure}[h]
		\begin{center}
			\includegraphics[width=0.42\textwidth]{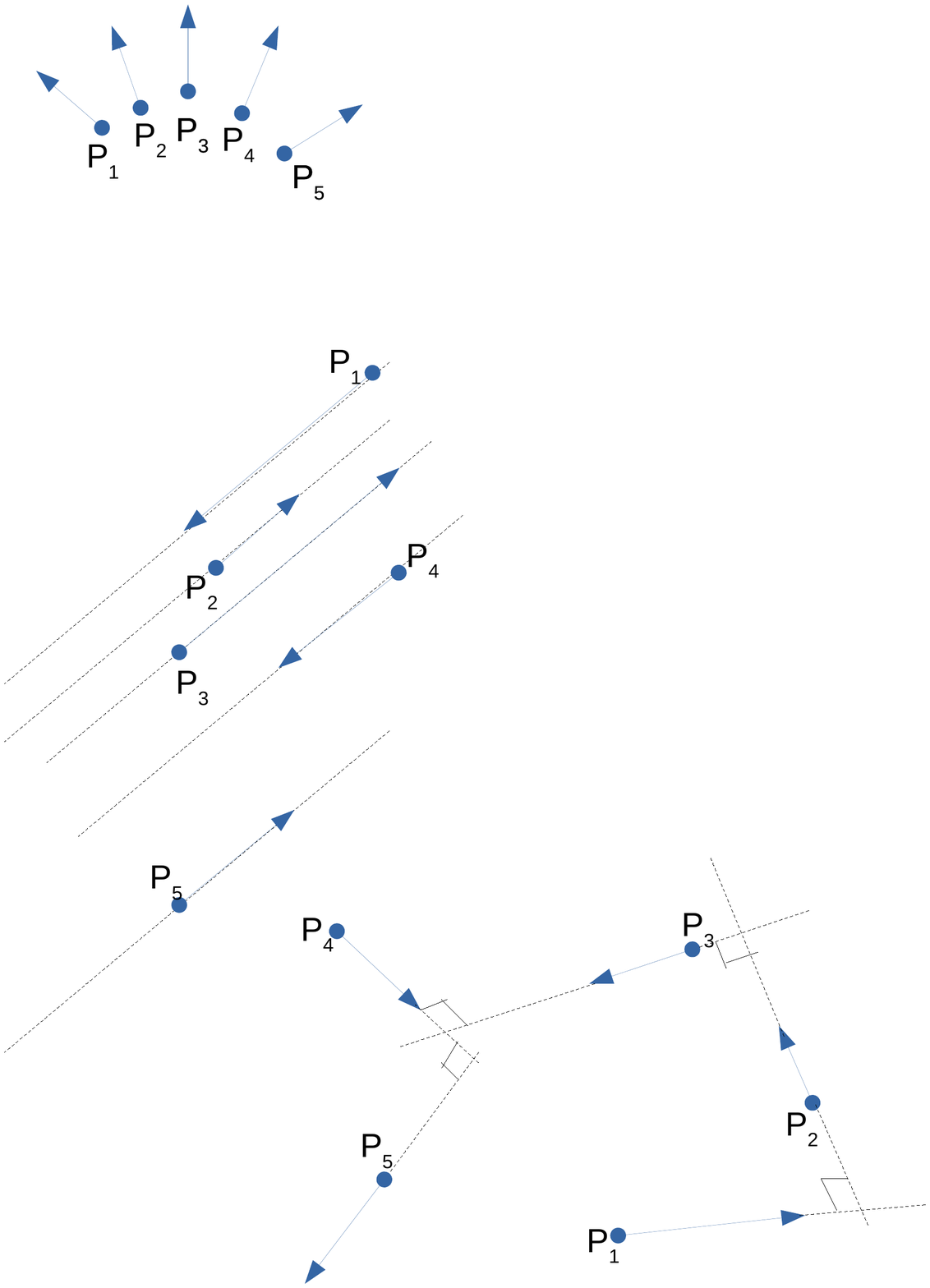}
			\caption{schematic representation of Example 2, top (same magnitude of velocities), Example 3, centre (parallel velocities) and Example 4 (perpendicular velocities), with $N=5$.}
		\end{center}
	\end{figure}
	
\begin{exe}
The circumstance of $N$ points with parallel velocities can be formulated by means of the restrictions
\begin{equation}
	\label{velparallvett}
{\dot P}_1\wedge {\dot P}_2={\dot P}_1\wedge {\dot P}_3=\,\dots \, ={\dot P}_1\wedge {\dot P}_N={\bf 0}
\end{equation}
which are equivalent to the $2(N-1)$ conditions

\begin{equation}
\label{velparall}
\left\{
\begin{array}{ll}
{\dot x}_1{\dot y}_2-{\dot x}_2{\dot y}_1=0, & {\dot x}_1{\dot z}_2-{\dot x}_2{\dot z}_1=0, \\
\qquad \dots &\qquad \dots \\
\qquad \dots &\qquad \dots \\
{\dot x}_1{\dot y}_N-{\dot x}_N{\dot y}_1=0, & {\dot x}_1{\dot z}_N-{\dot x}_N{\dot z}_1=0,
\end{array}
\right.
\end{equation}
Since  $n=3N$ and $k=2(N-2)$ we expect $m=3N-2(N-2)=N+2$: a natural choice for the $N+2$ independent velocities is 
\begin{equation}
\label{velparallind}
(\overbrace{{\dot q}_1, {\dot q}_2, \dots\; \dots, {\dot q}_N}^{N},
\overbrace{{\dot q}_{N+1}, {\dot q}_{N+2}}^{2})=(\overbrace{{\dot x}_1, {\dot x}_2, \dots, {\dot x}_N}^{N}, \overbrace{{\dot y}_1, {\dot z}_1}^{2})
\end{equation}
so that the explicit formulation (\ref{constrexpl}) for the $k=n-m=2(N-1)$ remaining velocities 
$({\dot q}_{N+3}, {\dot q}_{N+4}, \dots, {\dot q}_{3N})=
({\dot y}_2, {\dot y}_3, \dots, {\dot y}_N, {\dot z}_2, {\dot z}_3, \dots, {\dot z}_N)$ is
\begin{equation}
\label{velparallexpl}
\left\{
\begin{array}{ll}
{\dot y}_2={\dot q}_{N+3}=\alpha_1 =\dfrac{{\dot q}_{N+1}}{{\dot q_1}}{\dot q}_2 &
{\dot z}_2={\dot q}_{2N+2}=\alpha_N=\dfrac{{\dot q}_{N+2}}{{\dot q_1}}{\dot q}_2 \\
{\dot y}_3={\dot q}_{N+4}=\alpha_2 =\dfrac{{\dot q}_{N+1}}{{\dot q_1}}{\dot q}_3 &
{\dot z}_3={\dot q}_{2N+3}=\alpha_{N+1}=\dfrac{{\dot q}_{N+2}}{{\dot q_1}}{\dot q}_3 \\
	\quad \dots \qquad \dots & \quad \dots  \qquad \dots \\
{\dot y}_N={\dot q}_{2N+1}=\alpha_{N-1}=\dfrac{{\dot q}_{N+1}}{{\dot q_1}}{\dot q}_N  &
{\dot z}_N={\dot q}_{3N}=\alpha_{2N-2}=\dfrac{{\dot q}_{N+2}}{{\dot q_1}}{\dot q}_N
\end{array}
\right.
\end{equation}
\end{exe}
	
\begin{exe}
A different system consists in $N$ points in the space moving in a way that the velocity of each of them is perpendicular to the velocity of the previous one:
		\begin{equation}
			\label{velperpvett}
		{\dot P}_1\cdot {\dot P}_2={\dot P}_2\cdot {\dot P}_3=\dots = {\dot P}_{N-1}\cdot {\dot P}_N
		\end{equation}
The kinematic constraints (\ref{velperp}) are the $N-1$ conditions
	\begin{equation}
	\label{velperp}
\left\{
\begin{array}{l}
{\dot x}_1{\dot x}_2+{\dot y}_1{\dot y}_2 + {\dot z}_1 {\dot z}_2 = 0, \\
{\dot x}_2{\dot x}_3+{\dot y}_2{\dot y}_3 + {\dot z}_2 {\dot z}_3 = 0, \\
	\quad \dots \quad \dots \\
		{\dot x}_{N-2}{\dot x}_{N-1}+{\dot y}_{N-2}{\dot y}_{N-1} + {\dot z}_{N-2} {\dot z}_{N-1} = 0\\
	{\dot x}_{N-1}{\dot x}_N+{\dot y}_{N-1}{\dot y}_N + {\dot z}_{N-1} {\dot z}_N = 0
\end{array}
\right.
\end{equation}
The non--nullity of the velocities ${\dot P}_1$, ${\dot P}_2$, $\dots$,  ${\dot P_{N-1}}$ is a sufficient condition in order that he rank of these conditions is full, that is $N-1$.
With respect to (\ref{constr}) we have $n=3N$ (no holonomic constraint, we leave the cartesian notation instead of $q_1$, $\dots$, $q_n$ for the sake of simplicity), $r=k=N-1$ and $m=n-k=2N+1$. A suitable choice of the $2N+1$ independent velocities follows from the particular structure of the conditions (\ref{velperp}), which are combined in pairs: assuming at first $N$ odd, each of 
\begin{equation}
\label{perpair}
\left\{
\begin{array}{ll}
{\dot x}_{2j-1}{\dot x}_{2j}+{\dot y}_{2j-1}{\dot y}_{2j} + {\dot z}_{2j-1} {\dot z}_{2j} = 0, & \\
& j=1, \dots, \frac{1}{2}(N-1)\\
{\dot x}_{2j}{\dot x}_{2j+1}+{\dot y}_{2j}{\dot y}_{2j+1} + {\dot z}_{2j}{\dot z}_{2j+1}= 0, 
\end{array}
\right.
\end{equation}
can be solved whenever ${\dot P}_{2j-1}\wedge {\dot P}_{2j+1}\not ={\bf 0}$, taking as dependent 
two of the variables $({\dot x}_{2j}, {\dot y}_{2j}, {\dot z}_{2j})$: if the last two is the case, one has
\begin{equation}
\label{perpdip}
{\dot y}_{2j}={\dot x}_{2j}\dfrac{
{\dot z}_{2j-1}{\dot x}_{2j+1}-{\dot x}_{2j-1}{\dot z}_{2j+1}
}{{\dot y}_{2j-1}{\dot z}_{2j+1}-{\dot z}_{2j-1}{\dot y}_{2j+1}}, \qquad 
{\dot z}_{2j}={\dot x}_{2j}\dfrac{
{\dot x}_{2j-1}{\dot y}_{2j+1}-{\dot y}_{2j-1}{\dot x}_{2j+1}
}{{\dot y}_{2j-1}{\dot z}_{2j+1}-{\dot z}_{2j-1}{\dot y}_{2j+1}} \qquad j=1, \dots, \frac{1}{2}(N-1) 
\end{equation}
so that ${\dot x}_1, {\dot x}_2, \dots, {\dot x}_N$, ${\dot y}_1, {\dot y}_3, \dots, {\dot y}_N$, 
${\dot z}_1, {\dot z}_3, \dots, {\dot z}_N$ (odd index) are the $2N+1$ independent parameters expressing the $N-1$ variables 
${\dot y}_2, {\dot y}_4, \dots, {\dot y}_{N-1}$, ${\dot z}_2, {\dot z}_4, \dots, {\dot z}_{N-1}$ (even index). 

\noindent
If $N$ is even, equations (\ref{perpair}), (\ref{perpdip}) write the first $N-2$ conditions of (\ref{velperp}) for $j=1, \dots, \frac{N}{2}-1$, while the last equation provides one of the variables with index $N$, for instance
\begin{equation}
\label{lasteven}
{\dot z}_N=-\dfrac{1}{{\dot z}_{N-1}}({\dot x}_{N-1}{\dot x}_N + {\dot y}_{N-1}{\dot y}_N)
\end{equation}
so that ${\dot x}_1, {\dot x}_2, \dots, {\dot x}_N$, ${\dot y}_1, {\dot y}_3, \dots, {\dot y}_{N-1}$ (odd index), ${\dot y}_N$, ${\dot z}_1, {\dot z}_3, \dots, {\dot z}_{N-1}$ (odd index) are the $2N+1$ independent parameters expressing the $N-1$ variables ${\dot y}_2, {\dot y}_4, \dots, {\dot y}_{N-2}$, ${\dot z}_2, {\dot z}_4, \dots, {\dot z}_{N-2}, {\dot z}_N$ (even index). 

\noindent
When the additional condition of closure 
\begin{equation}
	\label{closurevett}
	{\dot P}_1\cdot {\dot P}_N=0
\end{equation}
is considered, then the constraint
\begin{equation}
	\label{closure}
	{\dot x}_1{\dot x}_N+{\dot y}_1{\dot y}_N + {\dot z}_1 {\dot z}_N=0
\end{equation}
has to be added to (\ref{velperp}). If $N$ is even (apart from the trivial case $N=2$, where  (\ref{velperp}) counts only one condition which is identical to (\ref{closure})), equation (\ref{lasteven}) combined with (\ref{closure}) gives
$$
{\dot y}_N = {\dot x}_N 
\dfrac{{\dot z}_1{\dot x}_{N-1}- {\dot x}_1{\dot z}_{N-1}}
{{\dot y}_1{\dot z}_{N-1}-{\dot y}_N{\dot z}_{N-1}} \qquad 
{\dot z}_N = {\dot x}_N 
\dfrac{{\dot y}_1{\dot x}_{N-1}- {\dot x}_1{\dot y}_{N-1}}
{{\dot y}_1{\dot z}_{N-1}-{\dot y}_N{\dot z}_{N-1}} 
$$
so that ${\dot y}_N$ switches to the group of dependent velocities ($k=N$, $m=2N$). 

\noindent
In the case of $N$ odd, the velocities in (\ref{closure}) are all independent and it suffices to make explicit one of them, such as
$$
{\dot z}_N=-\dfrac{1}{{\dot z}_1}({\dot x}_1{\dot x}_N + {\dot y}_1{\dot y}_N)
$$
which diminishes also in this case of one unity the number of independent velocities. 

\noindent
The procedure based on (\ref{perpdip}) is favored by the presence of a small number of independent velocities for each expression, however it fails when condition ${\dot P}_{2j-1}\wedge {\dot P}_{2j+1}\not ={\bf 0}$ is not valid: this occurs in the planar case, which is after all the most encountered case in literature (``nonholonomic chains''). As a matter of fact, in this case all the velocities ${\dot P}_i$ with even index are parallel, the same for the velocities with odd index; it follows that the closure condition (\ref{closure}) (`closed chain'') is automatically fulfilled for $N$ even, is unfeasible for $N$ odd (we still assume that the velocities are not null).
Assuming that $y=0$ is the plane which the points belong to, the constraints (\ref{velperp}) reduce to ${\dot x}_i{\dot x}_{i+1}+{\dot z}_i {\dot z}_{i+1}=0$ for $i=1, \dots, N-1$ and can be written explicitly by means of 
$$
\left\{
\begin{array}{l}
{\dot z}_2=-{\dot x}_2\dfrac{{\dot x}_1}{{\dot z}_1} \\
{\dot z}_3={\dot x}_3\dfrac{{\dot z}_1}{{\dot x}_1}\\ 
{\dot z}_4=-{\dot x}_4 \dfrac{{\dot x}_1}{{\dot z}_1}\\
\dots \quad \dots \quad \dots \\ 
{\dot z}_N=(-1)^{N+1}{\dot x}_N \left(\dfrac{{\dot x}_1}{{\dot z}_1}\right)^{(-1)^N}
\end{array}
\right.
$$
so that $k=N-1$ once again, $n=2N$ (actually the $N$ holonomic conditions $y_i=0$, $i=1 \dots, N$ are present) and the $m=n-k=N+1$ independent parameters are ${\dot x}_1$, ${\dot x}_2$, $\dots$, ${\dot x}_N$ ${\dot y}_1$. 
With respect to the general case (\ref{velperp}), it must be said that the analogous procedure in the space, that is achieving 
$$
{\dot z}_2=-\dfrac{1}{{\dot z}_1}({\dot x}_1{\dot x_2}+{\dot y}_1{\dot y}_2)
\quad \dots \quad \dots \quad
{\dot z}_N=-\dfrac{1}{{\dot z}_{N-1}}({\dot x}_{N-1}{\dot x_N}+{\dot y}_{N-1}{\dot y}_N)
$$
then expressing each of ${\dot z}_j$, $j=2, \dots, N$ in favour of the $2N+1$ independent velocities
by choosing ${\dot x}_1$, $\dots$, ${\dot x}_N$, ${\dot y}_1$, $\dots$, ${\dot y}_N$, ${\dot z}_N$, leads to very complex expressions presenting also problematic aspects when the case of closure (\ref{closure})
is contemplated.
\end{exe}
		
	\begin{exe} 
		Two points $P_1$ and $P_2$ are constrained on a
		plane $\pi$ and the straight lines orthogonal to the velocities ${\dot P}_1$, ${\dot P}_2$ intersect in a point of a given curve $\gamma$ lying on the plane. In other words, any point $P$ verifying $\overrightarrow{PP_1}\cdot {\dot P}_1=0$ and $\overrightarrow{PP_2}\cdot {\dot P}_2=0$ has to be a point of $\gamma$.
		
			\begin{figure}[h]
			\begin{center}
				\includegraphics[width=0.42\textwidth]{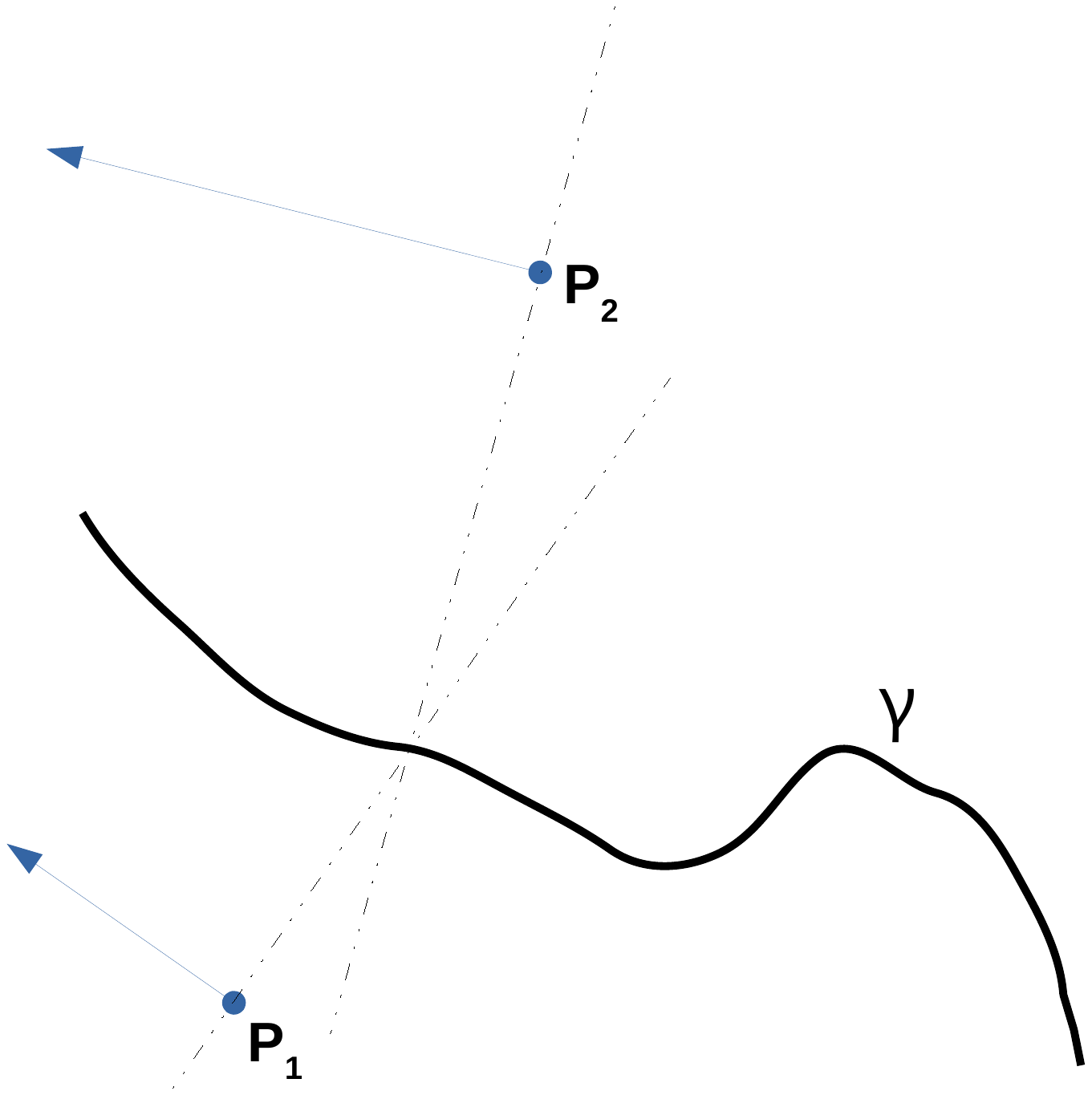}
				\caption{Graphic scheme of Example 5.}
			\end{center}
		\end{figure}
		
		\noindent
		Let us settle the cartesian frame of reference such that the plane $\pi$ is $z=0$: the equations of two straight lines are 
	\begin{equation}
		\label{dpxy}
		(x-x_1) {\dot x}_1+(y-y_1) {\dot y}_1=0, \qquad (x-x_2) {\dot x}_2+(y-y_2){\dot y}_2=0
	\end{equation}
where $(x_i,y_i,0)$ are the coordinates of $P_i$, $i=1,2$. The coordinates of the intersection  point,  definable if ${\dot P}_1\wedge {\dot P}_2\not ={\bf 0}$ that is ${\dot x}_1{\dot y}_2\not = {\dot x}_2{\dot y}_1$, are
$$
{\bar x}=
\dfrac{{\dot y}_1{\dot y}_2(y_1-y_2)-x_2{\dot y}_1 {\dot x}_2+x_1{\dot x}_1{\dot y}_2}
{{\dot x}_1{\dot y}_2 - {\dot x}_2{\dot y}_1},\;\;
{\bar y}=\dfrac{{\dot x}_1{\dot x}_2(x_2-x_1)-y_1{\dot y}_1 {\dot x}_2+y_2{\dot x}_1{\dot y}_2}
{{\dot x}_1{\dot y}_2 - {\dot x}_2{\dot y}_1}.
$$
Giving the curve $\gamma$ as the graph of $y=g(x)$, the kinematic constraint is then formulated in the following way:
$$
{\bar y}(x_1, y_1, x_2, y_2, {\dot x}_1, {\dot y}_1, {\dot x}_2, {\dot y}_2)=g({\bar x}(x_1, y_1, x_2, y_2, {\dot x}_1, {\dot y}_1, {\dot x}_2, {\dot y}_2))
$$
If $\gamma$ is the straight line $ax+by=0$, the constraint is
\begin{equation}
	\label{nonolpend}
a\left({\dot y}_1{\dot y}_2(y_1-y_2)-x_2{\dot y}_1 {\dot x}_2+x_1{\dot x}_1{\dot y}_2\right)+
b\left({\dot x}_1{\dot x}_2(x_2-x_1)-y_1{\dot y}_1 {\dot x}_2+y_2{\dot x}_1{\dot y}_2\right)=0.
\end{equation}
In particular, $\gamma$ as the $x$--axis (nonholonomic pendulum, studied in 
\cite{benenti}) yields $a=0$ and the kinematic condition
		\begin{equation}
		\label{nonolpenda0}
		{\dot x}_1(x_2{\dot x}_2+y_2{\dot y}_2)-{\dot x}_2(x_1{\dot x}_1+y_1{\dot y}_1)=0
		\end{equation}
Regarding the explicit form (\ref{constrexpl}), the two holonomic conditions of restriction on the plane make simply leave the two coordinates $z_1$ and $z_2$: defining (\ref{xqt})
		as $x_1=q_1$, $y_1=q_2$, $z_1=0$, $x_2=q_3$, $y_2=q_4$, $z_2=0$,  
the explicit form of (\ref{nonolpend}) is
 	\begin{equation}
 		\label{nhpexpl}
  	{\dot q}_4 = \dfrac{b(q_1-q_3){\dot q}_1+(bq_2+aq_3) {\dot q}_2}
 	{(aq_1+bq_4){\dot q}_1+a(q_2-q_4){\dot q}_2}{\dot q}_3 = \alpha_1 (q_1, q_2, q_3, q_4, {\dot q}_1, {\dot q}_2, {\dot q}_3)
 		\end{equation}
 	showing $n=4$ and $m=3$.
 	
	\end{exe}
	
	\noindent
	We point out that the constraints (\ref{samevel}), (\ref{velparall}), (\ref{velperp}), (\ref{nonolpend}) are represented by homogeneous quadratic kinematic functions. In terms of lagrangian coordinates and referring to (\ref{constr}), if the geometric constraints are absent or independent of time (see (\ref{xqt})) we can outline this category by   
		\begin{equation}
		\label{quadrom}
		\sum\limits_{i,j=1}^n a_{i,j}^{(\nu)}(q_1, \dots, q_n){\dot q}_i {\dot q}_j=0, \qquad \nu=1, \dots, k.
	\end{equation}
In particular, for (\ref{samevel}), (\ref{velparall}) and (\ref{velperp}) the coefficients $a_{i,j}^{(\nu)}$ are constant.

\subsection{Linear kinematic constraints}

\noindent
A special case which covers a wide framework of models and applications concerns the linear dependence of the kinematic conditions with respect to the velocities: this corresponds to set in (\ref{eqvinc}) 
$$
f_{h+1}({\bf X}, {\dot {\bf X}},t)={\bm \eta}_j({\bf X},t)\cdot {\dot {\bf X}}+\theta_j({\bf X}, t), \qquad j=1, \dots, k
$$
with ${\bm \eta}_j$ vector--valued function with values in ${\Bbb R}^{3N}$, $\theta_j$ real--value
function for each $j=1, \dots, k$.
In turn, the functions (\ref{fij}) are linear with respect to the kinetic variables ${\dot {\bf q}}$, so that (\ref{constr}) is the linear system

\begin{equation}
	\label{constrlin}
	\left\{
	\begin{array}{l}
		\sum\limits_{i=1}^n\sigma_{1,i}(q_1,\dots, q_n,t){\dot q}_i+\zeta_1(q_1, \dots, \dots, q_n, t)=0\\
		\qquad \qquad \dots  \dots  \dots\\
		\qquad \qquad \dots \dots \dots\\
		\sum\limits_{i=1}^n\sigma_{k,i}(q_1,\dots, q_n,t){\dot q}_i+\zeta_k(q_1, \dots, \dots, q_n, t)=0
		\end{array}
		\right.
\end{equation}
where
$$
\begin{array}{l}
	\sigma_{j,i}({\bf q},t)={\bm \eta}_j({\bf X}({\bf q},t),t)\cdot \dfrac{\partial {\bf X}}{\partial q_i}({\bf q},t), \quad j=1, \dots, k, \quad i=1, \dots, n \\ 
	\\
	\zeta_j({\bf q},t)={\bm \eta}_j({\bf X}({\bf q},t),t) \cdot \dfrac{\partial {\bf X}}{\partial t}+\theta_j({\bf X}({\bf q},t),t), \quad j=1, \dots, k
\end{array}	
$$
If the $k\times n$ matrix $(\sigma)_{\nu,i}$ has full rank $k$, then 
conditions (\ref{constrlin}) provide (\ref{constrexpl}) in the form
\begin{equation}
	\label{constrlinexpl}
\begin{cases}
	{\dot q}_{m+1}=\sum\limits_{j=1}^m \alpha_{1,j}(q_1, \dots, q_n, t){\dot q}_j+\beta_1(q_1, \dots, q_n,t) \\
	\dots  \\
	\dots  \\
	{\dot q}_n=\sum\limits_{j=1}^m \alpha_{k,j}(q_1, \dots, q_n, t){\dot q}_j+\beta_k(q_1, \dots, q_n,t)
\end{cases}
\end{equation}
for suitable coefficients $\alpha_{i,j}$ and $\beta_j$, $i=1, \dots, k$, $j=1, \dots, m$.

\subsection{Examples of linear nonholonomic systems}

\noindent
It is worth to mention that not a few examples of systems with linear constraints originate from  nonlinear conditions (e.~g.~parallelism, orthogonality, ..) by adding some specific request: 
an evident instance is the pair of constraints (\ref{dpxy}), which turn into linear if the intersection point $P\equiv (x,y)$ becomes part of the system.

\noindent
As a second instance, if in (\ref{velparallvett}) one specifies that the velocities are parallel to the same vector ${\bf v}=(\alpha, \beta, \gamma)$, the set (\ref{velparall}) is revised as
$$
\beta {\dot x}_i - \alpha {\dot y}_i=0, \quad \gamma {\dot x}_i - \alpha {\dot z}_i=0
$$
forming $2N$ linear kinematic conditions.

\noindent
Or else, if in (\ref{velperpvett}), $N=2$, it is required that the velocity of $P_2$ is also perpendicular to the straight line joining the two points, the kinematic constraints are
$$
{\dot P}_2 \cdot \overrightarrow{P_1P_2}=0, \quad {\dot P}_1 \wedge \overrightarrow{P_1P_2}={\bf 0}
$$
which are linear (we will discuss deeper the question in Example 8).

\noindent
For the most part, nonholonomic systems considered in literature deal with linear kinematic constraints: the rolling disc or the rolling sphere on a fixed or mobile surface are largely studied in texts and papers. The mentioned examples are also right for the question of introducing suitable pseudo--velocities (see among others \cite{bloch1}). Nevertheless, examples based on discrete points systems are more appropriate for our purposes and for being considered more times, in order to add further aspects (equations of motion, energy balance, ...).
At the same time, our intention is to make clear the set up of the selected examples hereafter, recurring in literature but sometimes lacking legibility about the independence among the exerted conditions.

\begin{exe}
A simple category of models taken as basic examples in most texts considers two points $P_1$ and $P_2$ moving on a plane and combines two (or more) of the following constraints:
\begin{itemize}
	\item[$(a)$]  the distance between the points is constant: $|\overrightarrow{P_1P_2}|=\ell>0$,
	\item[$(b)$]  the velocities have the same magnitude: $|{\dot P}_1|=|{\dot P}_2|$,
	\item[$(c)$]  the velocity of the midpoint $B$ of $\overrightarrow{P_1P_2}$ is parallel to the straight line joining the two points: ${\dot B}\wedge \overrightarrow{P_1P_2}={\bf 0}$.
\end{itemize}

\noindent
It is worth to make plain the inferences among them:
$$
\begin{array}{cl}
I &(a), \, (c)\;\Rightarrow\;(b) \\
II &(a), \, (b) \;\Rightarrow \;(c)\;\textrm{whenever}\;{\dot P}_1\not = {\dot P}_2 \\
III &(b), \, (c)\;\Rightarrow \;(a)\;\textrm{whenever}\;{\dot B}\not ={\bf 0}
\end{array}
$$
As a matter of fact, the formulation (\ref{eqvinc}) of the three constraints $(a)$, $(b)$ and $(c)$ is
$$
\left\{
\begin{array}{ll}
	(x_1-x_2)({\dot x}_1-{\dot x}_2)+(y_1-y_2)({\dot y}_1-{\dot y}_2)=0 & (a)\\
	\\
({\dot x}_1+{\dot x}_2)({\dot x}_1-{\dot x}_2)+({\dot y}_1+{\dot y}_2)({\dot y}_1-{\dot y}_2)=0
& (b)\\
	\\
(x_1-x_2)({\dot y}_1+{\dot y}_2)-(y_1-y_2)({\dot x}_1+{\dot x}_2)=0 & (c)
\end{array}
\right.
$$
As for statement $I$, conditions $(a)$, $(c)$ can be considered as a linear system for the two quantities $(x_1-x_2)$ and $(y_1-y_2)$; since $P_1\not \equiv P_2$, the existence of not null solutions entails that the determinant $-({\dot x}_1-{\dot x}_2)({\dot x}_1+{\dot x}_2)-({\dot y}_1-{\dot y}_2)({\dot y}_1+{\dot y}_2)$ vanishes, so that $(b)$ is valid.
For $II$ one argues analogously: $(a)$, $(b)$ is a system for $({\dot x}_1-{\dot x}_2)$ and $({\dot y}_1-{\dot y}_2)$; any not null solution (the null solution corresponds to ${\dot P}_1={\dot P}_2$) requires $(x_1-x_2)({\dot y}_1+{\dot y}_2)=(y_1-y_2)({\dot x}_1+{\dot x}_2)$, that is condition $(c)$. 
Lastly, $(b)$, $(c)$ is a system for ${\dot x}_1+{\dot x}_2$ and ${\dot y}_1+{\dot y}_2$; the null solution corresponds to ${\dot B}={\bf 0}$ and nullifying the determinant is equivalent to condition $(a)$.
\end{exe}

\begin{rem}
	Introducing standard coordinates for such as problems makes evident the geometrical or kinematical 
	meaning of the previous scheme. Actually, if $(a)$ is in effect, the holonomic constraint let us write (\ref{xqt}) in the form
	\begin{equation}
		\label{2pti}
		x_1=q_1+\dfrac{\ell}{2} \cos q_3, \;\;y_1=q_2+\dfrac{\ell}{2} \cos q_3, \;\;x_2=q_1-\dfrac{\ell}{2} \cos q_3,\;\;y_2=q_2-\dfrac{\ell}{2} \sin q_3
	\end{equation}
($q_1$ and $q_2$ are the coordinates of the midpoint, $q_3$ is the angle that $\overrightarrow{P_1P_2}$ forms with the $x$--axis, increasing anticlockwise). 
The form (\ref{constr}) in the lagrangian coordinates (\ref{2pti}) of the kinematic constraints is 
$$
\left\{
\begin{array}{ll}
{\dot q}_3({\dot q}_1 \sin q_3 - {\dot q}_2 \cos q_3)=0	& (b)\\
	\\
{\dot q}_1 \sin q_3 - {\dot q}_2 \cos q_3=0 & (c)
\end{array}
\right.
$$
which makes evident that if the velocity of the midpoint is along the conjoining line then the endpoints must have the same magnitude of the velocity (statement $I$); the opposite sense is not necessarily true, since in any translational motion (corresponding to ${\dot q}_3=0$) not parallel to $\overrightarrow{P_1P_2}$, the velocities of the ends are equal but ${\dot B}$ does not have the direction $\overrightarrow{P_1P_2}$. For this reason statement $II$ is valid for ${\dot P}_1\not = {\dot P}_2$, in such a way that a non--null angular velocity is present (${\dot q}_3\not =0$) and $(b)$, $(c)$ turn out to be equivalent. An anologous explication holds for statment $III$: if ${\dot B}$ is zero, a rotational motion around $B$ of the segment $P_1P_2$ shows the same magnitude of the velocities of the ends even if the lenght of the segment changes (so that $(a)$ fails).
\end{rem}

\begin{exe}
A second example with similar points concerns the following conditions:
\begin{itemize}
	\item[$(a)$] the distance between the points is constant: $|\overrightarrow{P_1P_2}|=\ell>0$,
	\item[$(b_1)$]  the velocities are parallel: ${\dot P}_1\wedge {\dot P}_2={\bf 0}$,
	\item[$(c_1)$]  the velocity of the midpoint $B$ is orthogonal to the straight line joining the two points: ${\dot B} \cdot \overrightarrow{P_1P_2}={\bf 0}$,
	\item[$(d_1)$] the velocities are orthogonal to the joining straight line: ${\dot P}_1\cdot \overrightarrow{P_1P_2}=0$, ${\dot P}_2\cdot \overrightarrow{P_1P_2}=0$
\end{itemize}

\noindent
We can prove the analogous statements:
$$
\begin{array}{ll}
I &	(d_1)\;\Rightarrow\;(a), \,(b_1),\, (c_1)\\
II &	(a), \, (c_1) \;\Rightarrow \;(b_1), \, (d_1) \\
III &	(b_1), \, (c_1) \;\Rightarrow \;(d_1)\;\textrm{whenever}\;{\dot B}\not = {\bf 0} \\
IV &	(a), \;	(b_1) \;\Rightarrow \;(d_1)\;\textrm{whenever}\;{\dot P}_1\not = {\dot P}_2 
\end{array}
$$

$$
\left\{
\begin{array}{ll}
	(x_1-x_2)({\dot x}_1-{\dot x}_2)+(y_1-y_2)({\dot y}_1-{\dot y}_2)=0& (a)\\
	\\
	{\dot x}_1{\dot y}_2 - {\dot x}_2{\dot y}_1=0 & (b_1)\\
	\\
	(x_1-x_2)({\dot x}_1+{\dot x}_2)+(y_1-y_2)({\dot y}_1+{\dot y}_2)=0 & (c_1)\\
	\\
		(x_1-x_2){\dot x}_1+(y_1-y_2){\dot y}_1=0, \;
		(x_1-x_2){\dot x}_2+(y_1-y_2){\dot y}_2=0 & (d_1)
\end{array}
\right.
$$
Property $I$ is evident: $(a)$ and $(c_1)$ are the sum and the difference of the conditions $(d_1)$, $(b_1)$ comes from the condition of existence of non--vanishing solutions of system $(d_1)$, which is linear with respect to $x_1-x_2$, $y_1-y_2$ (actually  $P_1\not \equiv P_2$). Arguing in the same way, system $(a)$, $(c_1)$ entails $(b_1)$ (as the condition of null determinant) and $(d_1)$ (by summing and subtracting), hence $II$ is valid. As for $III$, a simple calculation on $(b_1)$ and $(c_1)$ leads to
$({\dot x}_1+{\dot x}_2)[(x_1-x_2){\dot x}_1+(y_1-y_2){\dot y}_1]=0$,
$({\dot y}_1+{\dot y}_2)[(x_1-x_2){\dot x}_2+(y_1-y_2){\dot y}_2]=0$; it follows that, if ${\dot B}\not =0$, at least one of the two conditions $(d_1)$ (which appear in square brackets in the previous expressions) must be true and the other one follows from $(c_1)$. Finally, statement $IV$ is obtained in a similar way, by rearranging $(a)$ and $(b_1)$ and by assuming $({\dot x}_1-{\dot x}_2)^2+({\dot y}_1-{\dot y}_2)^2\not =0$ in order to get $(d_1)$. 

\noindent
We remark that in statement $III$ the assumption ${\dot B}\not = {\bf 0}$ is necessary, since in a rotational motion around $B$ the velocities of the ends are not orthogonal to the joining segment, if the lenght of the latter is not constant.
In the same way, statement $III$ fails in a translational motion along any direction not parallel to $\overrightarrow{P_1P_2}$: the critical case ${\dot P}_1={\dot P}_2$ allows the velocities to be not orthogonal to the joining line. 
\end{exe}

\begin{exe} 
Let us introduce a further example, examined in \cite{zek3}: two points move on a plane and
\begin{itemize}
	\item[$(a_2)$] the velocities are perpendicular: ${\dot P}_1\cdot {\dot P}_2=0$,
	\item[$(b_2)$]  the velocity of one of them is perpendicular to the straight line joining the points: ${\dot P}_1\cdot \overrightarrow{P_1P_2}=0$, 
	\item[$(c_2)$]  the velocity of the other point is parallel to the joining line: 
	${\dot P}_2\wedge \overrightarrow{P_1P_2}={\bf 0}$.
\end{itemize}

	\begin{figure}[h]
	\begin{center}
		\includegraphics[width=0.42\textwidth]{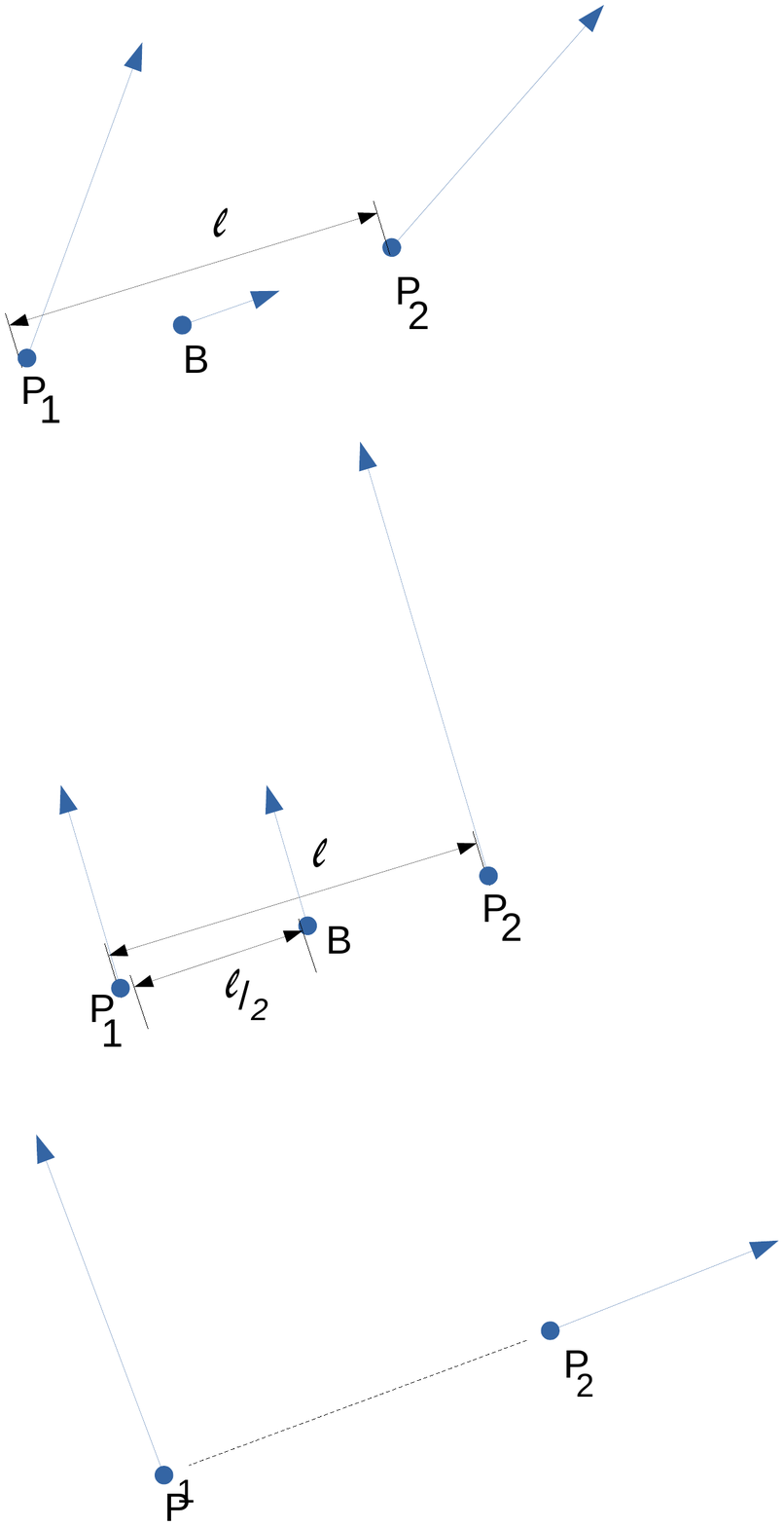}
		\caption{Schematic drawings of Example 6, 7 and 8 from the top to the bottom, respectively.}
	\end{center}
\end{figure}

\noindent
The setting (\ref{constr}) is 
$$
\left\{
\begin{array}{ll}
	{\dot x}_1{\dot x}_2+{\dot y}_1{\dot y}_2=0& (a_2)\\
	\\
	(x_1-x_2){\dot x}_1+(y_1-y_2){\dot y}_1=0 & (b_2)\\
	\\
	(x_1-x_2){\dot y}_2-(y_1-y_2){\dot x}_2=0 & (c_2)
\end{array}
\right.
$$
However, the three conditions are not independent: actually, it can be show that
$$
\begin{array}{ll}
	I &	(a_2), \, (b_2)\;\Rightarrow\; (c_2)\;\textrm{whenever}\;{\dot P}_1\not = {\bf 0} \\
	II &	(a_2), \, (c_2) \;\Rightarrow \;(b_2)\;\textrm{whenever} \; {\dot P}_2\not = {\bf 0} \\
	III &	(b_2), \, (c_2) \;\Rightarrow \;(a_2)\;\textrm{whenever} \;P_1\not \equiv P_2
\end{array}
$$
In order to prove that, it suffices to argue as in the previous examples; moreover, the critical configurations to exclude have an evident meaning.
\end{exe}

\begin{rem}
More than one of the examples we presented shows that in some cases the same system can be treated either with linear kinematic constraints or with nonlinear kinematic constraints (or a mixture of them): a crucial point is to know whether the two approaches are perfectly equivalent or they are incoherent in any critical configuration, as we highlighted in some case.
\end{rem}

\begin{exe}

\noindent
The following example concerns a linear kinematic constraint depending explicitly on time.
Let us consider in the three-dimensional space a reference point $Q\equiv (x_Q, y_Q, z_Q)$ moving according to the given relations $x_Q=x_Q(t)$, $y_Q=y_Q(t)$, $z_Q=z_Q(t)$. A point $P\equiv (x,y,z)$ ``pursues'' $Q$ in a way such that its velocity is at any time parallel to the straight line joining $P$ with $Q$: this means $({\dot x}, {\dot y}, {\dot z})\wedge (x_Q(t)-x, y_Q(t)-y,z_Q(t)-z) ={\bf 0}$ giving the two independent conditions 
\begin{equation}
	\label{constrpurs}
\left\{
\begin{array}{l}
(y_Q(t)-y){\dot x}-(x_Q(t)-x){\dot y}=0, \\
\\
(z_Q(t)-z){\dot x}-(x_Q(t)-x){\dot z}=0
\end{array}
\right.
\end{equation}

 which represent the constraints equations
(\ref{eqvinc}), both of kinematic type ($r=k=2$).

\noindent
Since no holonomic condition is present (that is $h=0$), the lagrangian coordinates are $q_1=x$, $q_2=y$, $q_3=z$ ($n=3$) and (\ref{xqt}) is ${\bf X}={\bf q}$. Moreover, $k=r=2$ (only kinematic constraints),  $m=1$ (one independent velocity) and the explicit form (\ref{constrlinexpl}) is, wherever $x\not= x_Q$, 
\begin{equation}
	\label{constrpursexpl}
\left\{
\begin{array}{l}
	{\dot q}_2 =\dfrac{y_Q(t)-q_2}{x_Q(t)-q_1}{\dot q}_1=\alpha_{1,1}(q_1, q_2, t){\dot q}_1, \\
	\\
	{\dot q}_3 =\dfrac{z_Q(t)-q_3}{x_Q(t)-q_1}{\dot q}_1=\alpha_{2,1}(q_1, q_3, t){\dot q}_1
\end{array}
\right.
\end{equation}

\noindent
In (\ref{constrpurs}) the point $P$ running after the moving object $Q$ can get around the whole space:
a modification can be done by considering the chasing point $P$ constrained to a regular surface $f(x,y,z)=0$ but the moving object $Q(t)$ not necessarily lying on the surface; the kinematic condition enforces the velocity to have the direction, at any time $t$, of the projection of $\overrightarrow{PQ(t)}$ on the tangent plane to the surface at $P$. In simple terms, $P$ runs after $Q(t)$ by choosing on the tangent plane the direction the closest to the joining straight line $PQ(t)$ as possible.

	\begin{figure}[h]
	\begin{center}
		\includegraphics[width=0.42\textwidth]{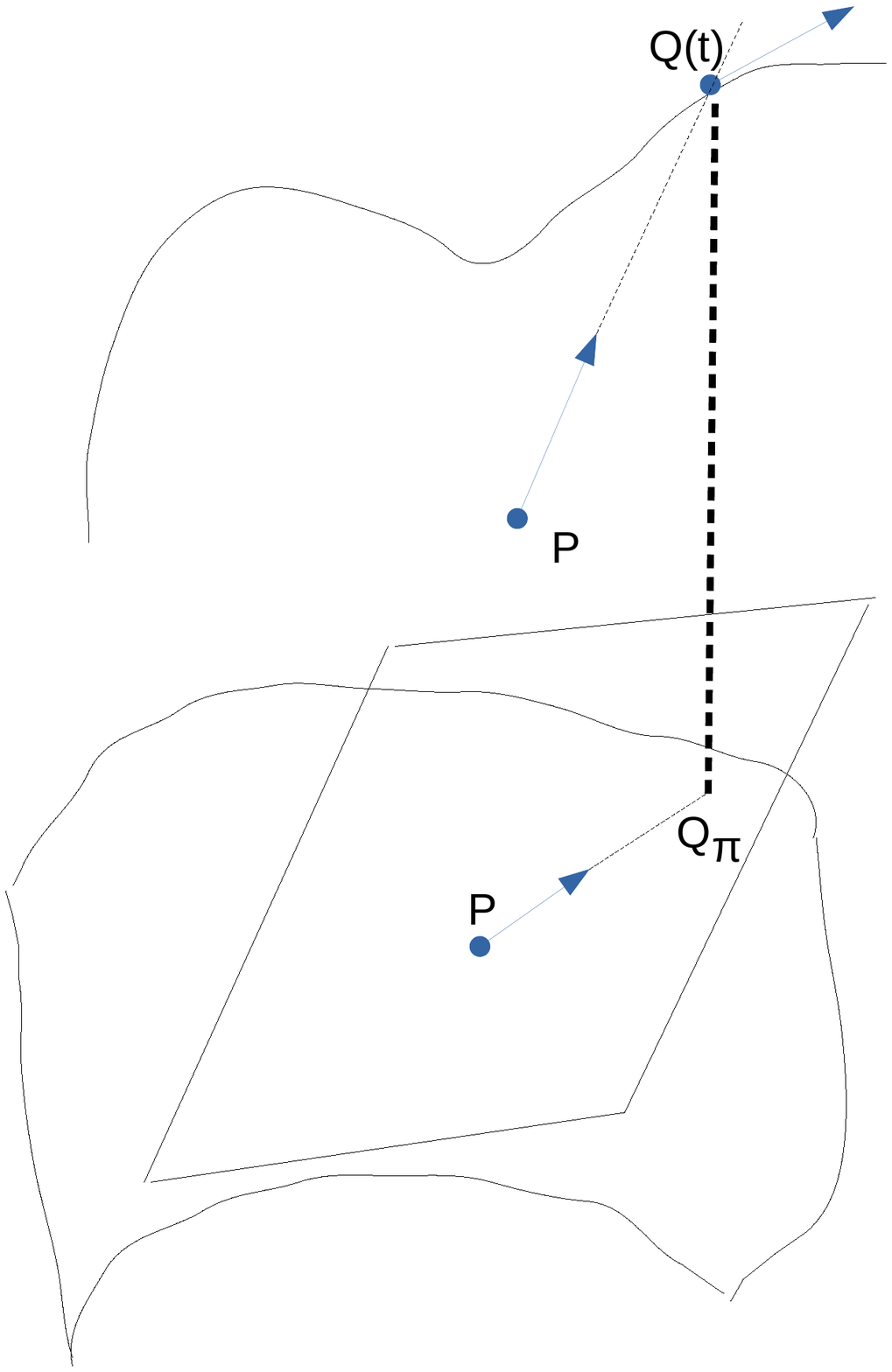}
		\caption{The pursuing model of Example 9, when $P$ is free in the space (above) and when $P$ is constrained on a surface (below.)}
	\end{center}
\end{figure}

\noindent
Calling $Q_\pi$ the orthogonal projection of $Q$ on the tangent plane to the surface at $P$, the kinematic constraint is 
\begin{equation}
\label{purssup}
{\dot P}\wedge \overrightarrow{PQ_\pi}={\bf 0}
\end{equation}
On the other hand, the vector $\overrightarrow{Q_\pi Q}$ is aligned with the direction $\left.\nabla_{\bf x}f\right\vert_{P}=
\left.(f_x, f_y, f_z)\right\vert_{P}$ perpendicular to the tangent plane at $P$: 
$$
\overrightarrow{Q_\pi Q}=\dfrac{\overrightarrow{PQ}\cdot \nabla_{\bf x}f}{|\nabla_{\bf x}f|^2}\nabla_{\bf x}f
$$
(it is assumed $|\nabla_{\bf x}f|\not =0$ anywhere). Having in mind that  $\overrightarrow{PQ_\pi}=\overrightarrow{PQ}-\overrightarrow{Q_\pi Q}$, 
the constraint (\ref{purssup}) is equivalent to the conditions 

\begin{equation}
	\label{purssupexpl}
	\left\{
	\begin{array}{l}
B(x,y,z,t){\dot x}-A(x,y,z,t){\dot y}=0, \\
\\
C(x,y,z,t){\dot x}-A(x,y,z,t){\dot z}=0	
\end{array}
\right.	
\end{equation}
where
$$
\begin{array}{l}
A(x,y,z,t)=(f_y^2+f_z^2)(x_Q(t)-x)-f_x(f_y(y_Q(t)-y)+f_z(z_Q(t)-z)), \\
\\
B(x,y,z,t)=(f_x^2+f_z^2)(y_Q(t)-y)-f_y(f_x(x_Q(t)-x)+f_z(z_Q(t)-z)), \\
\\
C(x,y,z,t)=(f_x^2+f_y^2)(z_Q(t)-z)-f_z(f_x(x_Q(t)-x)+f_y(y_Q(t)-y))
\end{array}
$$
being $(x,y,z)$, $(x_Q, y_Q, z_Q)$ the coordinates of respectively $P$ and $Q(t)$ and all the derivatives $f_x$, $f_y$, $f_z$ calculated at $P$.
The holonomic constraint $f(x,y,z)=0$ appears according to (\ref{eqvinc}) as 
$$
f_x{\dot x}+f_y {\dot y}+f_z {\dot z}=0.
$$
The latter condition and the two conditions (\ref{purssupexpl}) are actually not independent: as a matter of fact, the rank of (\ref{rankr}) is not full, according to 
$$
det\;\left(\begin{array}{ccc} C & 0 & -A\\ B & -A & 0 \\ f_x & f_y & f_z
\end{array}
\right)
=-A(Af_x+Bf_y+Cf_z)=0
$$
since the quantity in round brackets is null (indeed $(A,B,C)$ is parallel to $({\dot x}, {\dot y}, {\dot z})$), hence perpendicular to $(f_x, f_y, f_z)$).

\noindent
Hence, the problem can be treated as $h=1$ making explicit $f(x,y,z)=0$ by introducinf two lagrangian parameters $q_1$, $q_2$, so that (\ref{xqt}) is $x=x(q_1, q_2)$, $y=y(q_1, q_2)$, $z=z(q_1,q_2)$. 
Moreover, only one of (\ref{purssupexpl}) is independent ($k=1$): the explicit expression (\ref{constrlinexpl}) for ${\dot q}_2$ is achieved from one of the two kinematic constraints and writing (\ref{vellagr}) as ${\dot x}=x_{q_1}{\dot q}_1+x_{q_2}{\dot q}_2$ and similarly for ${\dot y}$, ${\dot z}$.

\noindent
A special case recurring in literature considers the flat surface $z=0$ and the trajectory $Q(t)$
belonging to the plane: the parametrization is simply $x=q_1$, $y=q_2$, $z=0$ and the kinematic conditions drastically reduce to $(y_Q-y){\dot x}-(x_Q-x){\dot y}=0$, ${\dot z}=0$ (see \cite{swac}).

\end{exe}

\begin{rem}
The constraints (\ref{constrlin}) ar linear affine functions of degree $1$. More broadly, for a positive integer $p$ the set of conditions
\begin{equation}
\label{constraff}
	\sigma_i(q_1, \dots, q_n, t)+\sum\limits_{j=1}^n\zeta_{i,j}(q_1, \dots, q_n,t){\dot q}_j^p=0, \qquad i=1, \dots, k
\end{equation}
refers to affine nonholonomic constraints of degree $p$. The explicit form (\ref{constrexpl}) corresponding to (\ref{constrlinexpl}) when $p=1$ is
\begin{equation}
	\label{constrexplaff}
{\dot q}_{m+\nu}^p=(\pm 1)^{p+1}\left(\sum\limits_{j=1}^m \alpha_{\nu,j}(q_1, \dots, q_n,t){\dot q}_j^p+\beta_\nu(q_1, \dots, q_n,t)\right)^{1/p}, \qquad \nu=1, \dots, k
\end{equation}
for suitable coefficients $\alpha_{\nu,j}$ and $\beta_\nu$.
The special case (\ref{pct}) can be seen as an affine constraint of degree $p=2$.	
	
\end{rem}

\section{The equations of motion}

\noindent
We now come back to (\ref{newton}), having in mind to write explicitly the right equations of motion.

\noindent
The equations of motion will be deduced from the d'Alembert principle, which is stated in ${\Bbb R}^{3N}$ by writing
\begin{equation}
	\label{dalamb}
	\left( {\dot {\bf Q}}-{\bf F}-{\bm \Phi}\right)	\cdot {\widehat {\dot {\bf X}}}={\bf 0}
\end{equation}
where
\begin{itemize}
	\item[] ${\bf Q}$ is the representative vector in ${\Bbb R}^{3N}$ of linear momenta: 
\begin{equation}
	\label{xm}
{\bf Q}={\dot {\bf X}}^{(\mathsf{M})}, \qquad {\bf X}^{(\mathsf{M})}=(M_1x_1, M_1y_1, M_1, z_1, \dots, M_N x_N, M_N y_N, M_N z_N)
\end{equation}
and $M_i$ is the mass of the $i$--th point, $i=1, \dots, N$, 
\item[] ${\bf F}=({\bf F}_1, \dots, {\bf F}_N)\in {\Bbb R}^{3N}$ makes the list of the active forces, being ${\bf F}_i$ the force acting on the $i$-th particle,
\item[] ${\bm \Phi}=({\bm \Phi}_1, \dots, {\bm \Phi}_N)\in {\Bbb R}^{3N}$ assembles  the vectors of the constraint forces, respectively, which give rise to the constraint conditions (\ref{eqvinc}),
\item[] ${\widehat {\dot {\bf X}}}\in {\Bbb R}^{3N}$ is any displacement consistent with the instantaneous configuration of the system, that is  at a blocked time $t$.
\end{itemize}

\noindent
The principle (\ref{dalamb}) is completed by the assumption of ideal constraints, that is demanding 
\begin{equation}
	\label{idconstr}
	{\bm \Phi}\cdot {\widehat {\dot {\bf X}}}=0   \qquad  \textrm{for any instanteneous compatible displacement}\;\;  {\hat {\dot {\bf X}}}.
\end{equation}

\noindent
Following this perspective, the main point is to delineate the right set of vectors ${\widehat {\dot {\bf X}}}$ compatible with the constraints (\ref{eqvinc}): if on one hand such a procedure is clear and recurring in the literature devoted to holonomic systems or systems with linear kinematic constraints, 
on the other the general case regarding nonlinearity (in itself rarely present) is commonly faced by different methods.

\noindent
Having in mind  the velocity representative vectors (\ref{vellagr}), (\ref{vellagrm}) and the explicit  expressions (\ref{constrexpl}) and (\ref{constrlinexpl}), we summarize in the following scheme the different sets of compatible instantaneous displacements
${\widehat {\dot {\bf X}}}$:
$$
 \begin{tabular}{c|c|c|}
 	&  ${\widehat {\dot {\bf X}}}$ & \\
 	 \hline \hline
\vspace{.2truemm}
Holonomic Constraints (HC)	& 
\vspace{.2truemm}
$\sum\limits_{j=1}^n  {\dot q}_j
\dfrac{\partial {\bf X}}{\partial q_j}$	& 
\vspace{.2truemm}
for arbitrary ${\dot q}_1, \dots, {\dot q}_n$\\
 \hline 
Linear Nonholonomic Constraints 	(LNC)	&	
\vspace{.5truemm}
$
\sum\limits_{r=1}^m  {\dot q}_r\left(
\dfrac{\partial {\bf X}}{\partial q_r}+
\sum\limits_{\nu=1}^{k}\alpha_{\nu,r}
\dfrac{\partial {\bf X}}{\partial q_{m+\nu}}\right)$
& 
\vspace{.1truemm}
for arbitrary ${\dot q}_1, \dots, {\dot q}_m$ \\
	\hline
Nonlinear Nonholonomic Constraints (NNC)
&	
$\sum\limits_{r=1}^m  {\dot q}_r\left(
\dfrac{\partial {\bf X}}{\partial q_r}+
\sum\limits_{\nu=1}^{k}\dfrac{\partial \alpha_\nu}{\partial {\dot q}_r}
\dfrac{\partial {\bf X}}{\partial q_{m+\nu}}\right)$ & 
for arbitrary ${\dot q}_1, \dots, {\dot q}_m$ \\
	\hline
\end{tabular}
$$
Let us add some comments. 
In the case of holonomic systems it is known and clear that the possible displacements compatible with instant constraint configuration are the elements of the linear space generated by the vectors $\frac{\partial {\bf X}}{\partial q_j}$, $j=1, \dots, n$, 
that is the tangent space.
In the case of linear constraints it is enough to take into account the expression (\ref{vellagrm}) to realize that the velocities in question form a linear subspace with respect to the holonomic case, since the relations (\ref{constrlinexpl}) are linear: eliminating the term caused by any movement of the constraint (which cannot take part in the evaluation of the movements allowed by the restrictions themselves) we therefore find in the (LNC) case the expression written in the box corresponding to ${\widehat {\bf {\dot X}}}$; the arbitrariness of the coefficients ${\dot q}_1$, $\dots$, ${\dot q}_m$ covers exactly the totality of the allowed displacements.

\noindent
Finally, for (NNC) systems lt us start from the observation that in both cases (HC) and (LNC) the following identity holds:
$$
{\widehat {\dot {\bf X}}}=\sum\limits_{r=1}^m
\dfrac{\partial {\dot {\bf X}}}{\partial {\dot q}_r}{\dot q}_r
$$
(the holonomic case implies $m=n$). The working hypothesis is to assume the relationship just written as the definition of the displacements admitted by the constraints even in the nonlinear case: 
taking into account (\ref{vellagrm}) we obtain
\begin{equation}
	\label{hatxnnc}
{\widehat {\dot {\bf X}}}=\sum\limits_{r=1}^m
\dfrac{\partial {\dot {\bf X}}}{\partial {\dot q}_r}{\dot q}_r=
\left(
\dfrac{\partial {\bf X}}{\partial q_r}+
\sum\limits_{\nu=1}^k \dfrac{\partial {\bf X}}{\partial q_{m+\nu}}
\dfrac{\partial \alpha_\nu}{\partial {\dot q}_r}\right){\dot q}_r
\end{equation}
which exactly coincides with the expression declared in the previous table.
\begin{rem}
The formula (\ref{hatxnnc}) presents itself in an advantageous way in order to write the equations of motion, since the arbitrariness of the factors  ${\dot q}_1, \dots, {\dot q}_m$ in ${\Bbb R}^m$ allows us to outline the possible directions in the vectors enclosed by the round brackets.
It is significant to compare the expression that would be obtained if the set of allowed displacements were considered as coming from (\ref{vellagrm}) ignoring the term due to the mobility of the constraint:
it should be written
$$
{\widehat {\dot {\bf X}}}=\sum\limits_{r=1}^m
\dfrac{\partial {\bf X}}{\partial q_r}{{\dot q}_r}+
\sum\limits_{\nu=1}^k \dfrac{\partial {\bf X}}{\partial q_{m+\nu}}
\dfrac{\partial \alpha_\nu}{\partial {\dot q}_r}
$$
which however does not offer the possibility, in varying the quantities ${\dot q}_r$, of identifying the directions along which to project the equations of motion.
It is interesting to observe that the two expressions coincide whenever the following hypothesis holds:
\begin{equation}
	\label{baralphaalpha}
\sum\limits_{i=1}^m {\dot q}_i\dfrac{\partial \alpha_\nu}{\partial {\dot q}_i}=\alpha_\nu\quad \textrm{for each} \quad \nu=1,\dots, k 	
\end{equation}\
which is definitely verified if $\alpha_\nu$ is a homogeneous function of degree one with respect to
kinetic variables ${\dot q}_1$, $\dots$, ${\dot q}_m$.

\end{rem}

\noindent
The corresponding equations of motions come from (\ref{dalamb}), seeped through (\ref{idconstr}) and the arbitrariness of the displacements pertaining each case (we also recall the completing equations (\ref{constrexpl}) and (\ref{constrlinexpl}) in the case (LNC) and (NNC), respectively):

\begin{equation}
	\label{hlnc}
\begin{array}{ll}
(HC) & 	\left( {\dot {\bf Q}}-{\bf F}\right)\cdot \dfrac{\partial {\bf X}}{\partial q_j}=0, \; j=1, \dots, n\\
	 & 	\left( {\dot {\bf Q}}-{\bf F}\right)\cdot {\bf X}_r=0, \; r=1, \dots, m
\left\{
\begin{array}{ll}
(LNC) & {\bf X}_r = \dfrac{\partial {\bf X}}{\partial q_r}+
\sum\limits_{\nu=1}^{k}\alpha_{\nu,r}\dfrac{\partial {\bf X}}{\partial q_{m+\nu}}\\
\\
& 
\textrm{jointly with}\;{\dot q}_\nu =\sum\limits_{j=1}^ m \alpha_{\nu,j}{\dot q}_j+\beta_\nu, \; \nu=1, \dots, k  \\ 	
\\ 
\\
(NNC) & {\bf X}_r=\dfrac{\partial {\bf X}}{\partial q_r}+
\sum\limits_{\nu=1}^{k}\dfrac{\partial \alpha_\nu}{\partial {\dot q}_r}
\dfrac{\partial {\bf X}}{\partial q_{m+\nu}}
\\   
\\
& 
\textrm{jointly with}\; {\dot q}_\nu = \alpha_\nu, \; 
\nu=1, \dots, k
\end{array}
\right.
\end{array}
\end{equation}
In all three cases (HC), (LNC) and (NNC) the system enumerates $n$ equations in the $n$ unknown functions $(q_1, \dots, q_n)$. In the nonholonomic cases (LNC) and (NNC), 
the $m$ equations of motion (\ref{dalamb}) have to be joined to the $k$ equations (\ref{constrexpl}) in order to form a differential system of $m+k=n$ equations. 

\subsection{The equations of motions in the lagrangian form}

\noindent
Let us introduce the kinetic energy of the system  $T=\sum\limits_{i=1}^N m_i {\dot {\bf x}_i^2}$, which can be written by means of the ${\Bbb R}^{3N}$--representative vectors (\ref{xm}) as
\begin{equation}
	\label{encindef}
T=\dfrac{1}{2}{\bf Q}\cdot {\dot {\bf X}}. 
\end{equation}
In a (HC) system one has 
$T=T(q_1, \dots, q_n, {\dot q}_1, \dots, {\dot q}_n, t)$, where the dependence on the listed variables is  acquired via (\ref{vellagr}).

\noindent
The well known property
$$
{\dot {\bf Q}}\cdot \dfrac{\partial {\bf X}}{\partial q_j}=
\dfrac{d}{dt}\dfrac{\partial T}{\partial {\dot q}_j}-\dfrac{\partial T}{\partial q_j}, 
\qquad j=1, \dots, n
$$
is sufficient to writing the equations of motion of a (HC) system in terms of the kinetic energy.
The same property can be employed for nonholonomic systems: let us look after (NNC) systems and comment later the linear case (LNC)
and write
\begin{equation}
	\label{qxr}
{\dot {\bf Q}}\cdot {\bf X}_r={\dot {\bf Q}}\cdot \dfrac{\partial {\bf X}}{\partial q_r}+
\sum\limits_{\nu=1}^{k}\dfrac{\partial \alpha_\nu}{\partial {\dot q}_r} {\dot {\bf Q}}\cdot 
\dfrac{\partial {\bf X}}{\partial q_{m+\nu}}=
\dfrac{d}{dt}\dfrac{\partial T}{\partial {\dot q}_r}-\dfrac{\partial T}{\partial q_r}+ 
\sum\limits_{\nu=1}^k 
\dfrac{\partial \alpha_\nu}{\partial {\dot q}_r}\left(
\dfrac{d}{dt}\dfrac{\partial T}{\partial {\dot q}_{m+\nu}}-\dfrac{\partial T}{\partial q_{m+\nu}}\right)
\end{equation}
for each $r=1, \dots, m$. At this point it is essential to refer to the independent kinetic variables: we then define
\begin{equation}
	\label{trid}
	\begin{array}{l}
	T^*(q_1,\dots, q_n, {\dot q}_1, \dots, {\dot q}_m, t)\\
	\\
	=T(q_1, \dots, q_n, {\dot q}_1, \dots, {\dot q}_m, \alpha_1(q_1, \dots, q_n, {\dot q}_1,\dots, {\dot q}_n,t), \dots, 
	\alpha_k(q_1, \dots, q_n, {\dot q}_1,\dots, {\dot q}_n,t), t)
	\end{array}
\end{equation}
as the kinetic energy restricted to the independent kinetic parameters $({\dot q}_1, \dots, {\dot q}_m)$. In (\ref{trid}) the functions $\alpha_\nu$ are the same as in (\ref{constr}) (possibly linear in the case (\ref{constrlinexpl}).
\begin{rem}
Recalling (\ref{vellagr}), the explicit expression of (\ref{encindef}) is 
\begin{equation}
	\label{encin}
T(q_1, \dots, q_n, {\dot q}_1, \dots, {\dot q}_n,t)=\dfrac{1}{2}
\sum\limits_{i,j=1}^n g_{i,j}{\dot q}_i {\dot q}_j +\sum\limits_{i=1}^n b_i {\dot q}_i+c
\end{equation}
where 
\begin{equation}
	\label{gbc}
g_{i,j}(q_1,\dots, q_n, t)=\dfrac{\partial {\bf X}^{(\mathsf{M})}}{\partial q_i}\cdot 
\dfrac{\partial {\bf X}}{\partial q_j}, \;\;
b_i(q_1,\dots, q_n,t)=\dfrac{\partial {\bf X}^{(\mathsf{M})}}{\partial q_i}\cdot \dfrac{\partial {\bf X}}{\partial t}, \;\;
c(q_1,\dots, q_n,t)=\frac{1}{2}
\dfrac{\partial {\bf X}^{(\mathsf{M})}}{\partial t}\cdot 
\dfrac{\partial {\bf X}}{\partial t}
\end{equation}
 so that (\ref{trid}) writes
\begin{equation}
	\label{tmobile}
	T^*=
	\frac{1}{2}\left( \sum\limits_{r,s=1}^m g_{r,s}{\dot q}_r {\dot q}_s+
	\sum\limits_{\nu, \mu=1}^k g_{m+\nu,m+\mu} \alpha_\nu \alpha_\mu \right)+
	\sum\limits_{r=1}^m \sum\limits_{\nu=1}^k g_{r,m+\nu}{\dot q}_r \alpha_\nu+
	\sum\limits_{r=1}^m b_r {\dot q}_r +\sum\limits_{\nu=1}^k b_{m+\nu}\alpha_\nu+c
\end{equation}

\end{rem}

\noindent
By virtue of the interrelations (easily inferable) among the derivatives of $T$ and $T^*$
\begin{equation}
	\label{relder}
\dfrac{\partial T}{\partial {\dot q}_r}=\dfrac{\partial T^*}{\partial {\dot q}_r}- 
\sum\limits_{\nu=1}^k \dfrac{\partial \alpha_\nu}{\partial {\dot q}_r}
\dfrac{\partial T}{\partial {\dot q}_{m+\nu}}, \;\;r=1, \dots, m \qquad
\dfrac{\partial T}{\partial q_j}=\dfrac{\partial T^*}{\partial q_j}- 
\sum\limits_{\nu=1}^k \dfrac{\partial \alpha_\nu}{\partial q_j}
\dfrac{\partial T}{\partial {\dot q}_{m+\nu}}, \;\;j=1, \dots, n \qquad
\end{equation}
it is possible to express (\ref{qxr}) in terms of $T^*$:
\begin{equation}
\label{tqxr}
\begin{array}{l}
{\dot {\bf Q}}\cdot {\bf X}_r=
\dfrac{d}{dt}\dfrac{\partial T^*}{\partial {\dot q}_r}-\dfrac{\partial T^*}{\partial q_r}
-\sum\limits_{\nu=1}^k\dfrac{\partial T^*}{\partial q_{m+\nu}}\dfrac{\partial \alpha_\nu}{\partial {\dot q_r}}
-\sum\limits_{\nu=1}^k   \dfrac{\partial T}{\partial {\dot q}_{m+\nu}}\left(
\dfrac{d}{dt}\left( \dfrac{\partial \alpha_\nu}{\partial {\dot q}_r}\right)- 
\dfrac{\partial \alpha_\nu}{\partial q_r}-\sum\limits_{\mu=1}^k 
\dfrac{\partial \alpha_\mu}{\partial {\dot q}_r}
\dfrac{\partial \alpha_\nu}{\partial q_{m+\mu}}\right)
\end{array}
\end{equation}
At this point we can state the following 
\begin{prop}
The equations of motion (\ref{hlnc}) for (NNC) systems subject to the kinematic constraints (\ref{constrexpl}) are 

\begin{equation}
	\label{vnl}
	\dfrac{d}{dt}\dfrac{\partial T^*}{\partial {\dot q}_r}-\dfrac{\partial T^*}{\partial q_r}
	-\sum\limits_{\nu=1}^k\dfrac{\partial T^*}{\partial q_{m+\nu}}\dfrac{\partial \alpha_\nu}{\partial {\dot q_r}}
	-\sum\limits_{\nu=1}^k  B_r^\nu \dfrac{\partial T}{\partial {\dot q}_{m+\nu}}
	={\cal F}^{(q_r)}+\sum\limits_{\nu=1}^k \dfrac{\partial \alpha_\nu}{\partial {\dot q}_r} {\cal F}^{(q_{m+\nu})}, 
	\qquad r=1,\dots, m
\end{equation}
where the coefficients $B_r^{\nu}$ are calculated by
\begin{equation}
	\label{b}
	B_r^\nu(q_1,\dots, q_n, {\dot q}_1, \dots, {\dot q}_m,t)= \dfrac{d}{dt}\left( \dfrac{\partial \alpha_\nu}{\partial {\dot q}_r}\right)- 
	\dfrac{\partial \alpha_\nu}{\partial q_r}-\sum\limits_{\mu=1}^k 
	\dfrac{\partial \alpha_\mu}{\partial {\dot q}_r}
	\dfrac{\partial \alpha_\nu}{\partial q_{m+\mu}}
\end{equation}
and 
\begin{equation}
\label{fqr}
{\cal F}^{(q_j)}={\bf F}\cdot \dfrac{\partial {\bf X}}{\partial q_j}, \quad j=1, \dots, n.
\end{equation}
\end{prop}

\noindent
The variables $({\dot q}_{k+1}, \dots, {\dot q}_n)$ appearing in the function $\dfrac{\partial T}{\partial {\dot q}_{m+\nu}}$ of (\ref{vnl}) must be expressed in terms of $(q_1,\dots, q_n,$ ${\dot q}_1, \dots, {\dot q}_m,t)$ by means of (\ref{constrexpl}). The same care has to be adopted for the explicit calculation of (\ref{b}):
	\begin{equation}
	\label{b2}
	B_r^{\nu}
	=\sum\limits_{j=1}^m
	\left( \dfrac{\partial^2 \alpha_\nu}{\partial {\dot q}_r \partial q_j}{\dot q}_j +
	\dfrac{\partial^2 \alpha_\nu}{\partial {\dot q}_r \partial {\dot q}_j}{\q2dot^{..}}_j\right)
	-\dfrac{\partial \alpha_\nu}{\partial q_r}
	+\sum\limits_{\mu=1}^k\left( \dfrac{\partial^2 \alpha_\nu}
	{\partial {\dot q}_r \partial q_{m+\mu}}\alpha_\mu-
	\dfrac{\partial \alpha_\mu}{\partial {\dot q}_r}\dfrac{\partial \alpha_\nu}{\partial q_{m+\mu}}\right)+
	\dfrac{\partial^2 \alpha_\nu}{\partial {\dot q}_r \partial t}.
\end{equation}

\subsection{Some remarks on the equations of motion}

\noindent
The equations of motion (\ref{vnl}) extend to the nonlinear case the Voronec equations appeared in \cite{voronec} and more recently discussed in \cite{neimark} for linear nonholonomic constraints (\ref{constrlinexpl}). The main connotation of the equations of this typology consists in using the real velocities of the system.

\noindent
The possibility of applying equations (\ref{vnl}) is very vast and they only require the explicit writing of the constraint conditions; the only limitation concerns the use of the real variables without involving the pseudo coordinates. In (\cite{zek2}) the same goal of deriving the more general form of the equations of motion is pursued in , using more generally the pseudo--coordinates (Poincar\'e--Chetaev variables); actually, when the latter coincide with the real generalized velocities, the equations are the same as we wrote.
Also in other cases in which the geometric approach is used for the description of nonlinear nonholonomic mechanical systems (Lagrangian systems on fibered manifolds) the final motion equations are the same as (\ref{vnl}).

\noindent
We examine below some aspects regarding the equations of motion and we point some special cases of (\ref{vnl}).

\subsubsection*{The linear case}
The equations for linear systems (LNC) are achieved by setting 
$\dfrac{\partial \alpha_\nu}{\partial {\dot q}_r}=\alpha_{\nu,r}$ in (\ref{vnl}) and the coefficients (\ref{b2}) as $B_r^\nu=\sum\limits_{r=1}^m 
b_{r,j}^\nu{\dot q}_j+\dfrac{\partial \alpha_{\nu,r}}{\partial t}$ with 
\begin{equation}
	\label{blin}
		b_{r,j}^\nu(q_1, \dots, q_n)=
		\dfrac{\partial \alpha_{\nu, r}}{\partial q_j}-
		\dfrac{\partial \alpha_{\nu,j}}{\partial q_r}+
		\sum\limits_{\mu=1}^k\left(
		\dfrac{\partial \alpha_{\nu, r}}{\partial q_{m+\mu}}\alpha_{\mu,j}-
		\dfrac{\partial \alpha_{\nu, j}}{\partial q_{m+\mu}}\alpha_{\mu,r}
		\right).
\end{equation}
so that equations (\ref{vnl}) take the form 
\begin{equation}
	\label{voronec}
	\dfrac{d}{dt}\dfrac{\partial T^*}{\partial {\dot q}_r}-\dfrac{\partial T^*}{\partial q_r}
	-\sum\limits_{\nu=1}^k\alpha_{\nu,r}\dfrac{\partial T^*}{\partial q_{m+\nu}}
	-\sum\limits_{\nu=1}^k\sum\limits_{j=1}^m 
	b_{r,j}^\nu{\dot q}_j\dfrac{\partial T}{\partial {\dot q}_{m+\nu}}=
	{\mathcal F}^{(q_r)}+\sum\limits_{\nu=1}^{k} \alpha_{\nu,r} {\cal F}^{(q_{m+\nu})}, \quad r=1, \dots, m
\end{equation}
known as Voronec's equations (\cite{neimark}).

\begin{exe}
We call to mind Example 8, concerning two points on a plane verifying ${\dot P}_1\cdot \overrightarrow{P_1P_2}=0$, ${\dot P}_2\wedge \overrightarrow{P_1P_2}={\bf 0}$ which correspond to the linear kinematic constraints $(b_2)$ and $(c_2)$ of the same Example. 
Setting (\ref{xqt}) as $x_1=q_1$, $y_1=q_3$, $x_2=q_2$, $y_2=_4$ the explicit formulation is 

$$
{\dot q}_3=-\dfrac{q_1-q_2}{q_3-q_4}{\dot q}_1, \qquad {\dot q}_4=\dfrac{q_3-q_4}{q_1-q_2}{\dot q_2}
$$
so that with respect to (\ref{constrlinexpl}) it is $k-2$ and $m=2$, $\alpha_{1,1}=-(q_1-q_2)/(q_3-q_4)$, $\alpha_{1,2}=\alpha_{2,1}=0$, $\alpha_{2,2}=(q_3-q_4)/(q_1-q_2)$, $\beta_1=\beta_2=0$. 
The function (\ref{trid}) is 
$$
T^*(q_1, q_2, q_3, q_4, {\dot q}_1, {\dot q}_2)=\dfrac{1}{2}M_1 \left(1+\alpha_{1,1}^2\right){\dot q}_1^2+\dfrac{1}{2}\left(1+\alpha_{2,2}^2\right){\dot q}^2_2
$$
where $M_1$ and $M_2$ are the masses. The calculation of the coefficients (\ref{blin}) leads to
$$
\begin{array}{ll}
b_{1,1}^1=b_{1,1}^2=b_{1,2}^1=0, \;\; b_{1,2}^2=\dfrac{1}{q_3-q_4}(1+\alpha_{2,2}^2)& \textrm{first equation}\;(r=1),\\
\\
b_{2,1}^1=b_{2,2}^1=b_{2,2}^2=0, \;\; b_{2,1}^2=-b_{1,2}^2& \textrm{second equation}\;(r=2).
\end{array}
$$
Assuming that the two points are connected by a spring exerting the force $-\kappa (P_1-P_2)$ on $P_1$ and the opposite one on $P_2$ ($\kappa$ positive constant) and including also the gravitational force directed in the direction of decreasing $y$, the equations of motion (\ref{voronec}) are
$$
\left\{
\begin{array}{l}
M_1\left(1+\alpha_{1,1}^2\right){\q2dot^{..}}_1+M_1\dfrac{q_1-q_2}{(q_3-q_4)^2}\left(1+\alpha_{1,1}^2
\right){\dot q}_1^2 =  -\kappa [(q_1-q_2)+\alpha_{1,1}(q_3-q_4)]-M_1\alpha_{1,1}g,\\
M_2\left(1+\alpha_{2,2}^2\right){\q2dot^{..}}_2-M_2\dfrac{1}{q_1-q_2}M_2
\left(1+\alpha_{2,2}^2\right){\dot q}_1^2 =
\kappa[(q_1-q_2)+\alpha_{2,2}(q_3-q_4)]-M_2\alpha_{2,2}g
\end{array}
\right.
$$
In \cite{zek3} the qualitative analysis of the model is extensively performed. 
\end{exe}

\begin{rem}
The nonholonomic device can be expanded by adding a third point $P_3$ (still on the same plane) which interplays with $P_2$ in the same way as the first pair of points, that is 
$$
{\dot P}_2\cdot \overrightarrow{P_2P_3}=0, \qquad {\dot P}_3\wedge \overrightarrow{P_2P_3}={\bf 0}.
$$
The construction can continue up to $N$ points, giving rise to the so--called nonholonomic chains (\cite{zek1}): the complete scheme of constraints is
\begin{equation}
	\label{ichain}
{\dot P}_i\cdot \overrightarrow{P_iP_{i+1}}=0, \qquad {\dot P}_{i+1}\wedge \overrightarrow{P_iP_{i+1}}={\bf 0}, \qquad i=1, \dots, N-1
\end{equation}
and the cartesian formulation (\ref{constr}) is given by the $2(N-1)$ conditions
$$
{\dot x}_i(x_i-x_{i+1})+{\dot y}_i (y_i - y_{i+1})=0, \qquad 
{\dot x}_{i+1}(y_i-y_{i+1})-{\dot y}_{i+1} (x_i - x_{i-1})=0, \qquad i=1, \dots, N-1
$$
Nevertheless, it must be said that the pair of constraints formed by the second of index $i$ and the first of index $i+1$, namely
$$
\left\{
\begin{array}{l}
{\dot x}_{i+1}(y_i-y_{i+1})-{\dot y}_{i+1} (x_i - x_{i-1})=0, \\
{\dot x}_{i+1}(x_{i+1}-x_{i+2})+{\dot y}_{i+1} (y_{i+1} - y_{i+2})=0
\end{array}
\right.
$$
entails the holonomic constraint $(x_i-x_{i+1})(x_{i+1}-x_{i+2})+(y_i-y_{i+1})(y_{i+1}-y_{i+2})=0$, expressing the orthogonality condition 
\begin{equation}
\label{olprob}
\overrightarrow{P_iP_{i+1}}\cdot \overrightarrow{P_{i+1}P_{i+2}}=0\;\;\textrm{for each}\; i=1, \dots, N-2.
\end{equation}
Hence, the problem requires to be set up with $N+2$ lagrangian coordinates which can be, as an instance, $(x_1, y_1)$, the angle $\varphi$ that $\overrightarrow{P_1P_2}$ forms with the horizontal direction and the $N-1$ abscissae $\xi_{i,i+1}$, $i=1, \dots, N-1$ on the straight lines $\overrightarrow{P_iP_{i+1}}$. The presence of the kinematic condition ${\dot P}_1\cdot \overrightarrow{P_1P_2}$ makes the problem different from the merely holonomic problem (\ref{olprob}): as a matter of facts, the kinematic 
restrictions generated by (\ref{olprob}) are not (\ref{ichain}), but 
$$
{\dot P}_{i+1}\cdot \overrightarrow{P_{i+1}P_{i+2}}-
{\dot P}_i\cdot \overrightarrow{P_{i+1}P_{i+2}}+
{\dot P}_{i+2}\cdot \overrightarrow{P_{i}P_{i+1}}-
{\dot P}_{i+1}\cdot \overrightarrow{P_{i}P_{i+1}}=0, \qquad i=1, \dots N-2
$$
\end{rem}

\subsubsection*{The stationary case} 
The situation corresponding to the independence of conditions (\ref{eqvinc}) from time $t$ explicitly (fixed or scleronomic constraints, either holonomic or nonholonomic) entails that $t$ is absent in (\ref{constr}), hence in $\alpha_\nu$, $\alpha_{\nu,j}$ and that  (\ref{tmobile}) is the form
$$
T^*=
\frac{1}{2}\left( \sum\limits_{r,s=1}^m g_{r,s}{\dot q}_r {\dot q}_s+
	\sum\limits_{\nu, \mu=1}^k g_{m+\nu,m+\mu} \alpha_\nu \alpha_\mu \right)+
		\sum\limits_{r=1}^m \sum\limits_{\nu=1}^k g_{r,m+\nu}{\dot q}_r \alpha_\nu
$$
(again, $t$ does not appear in $g_{r,s}$, $r,s=1, \dots, n$). The only difference in the equations 	
of motion is the lack in (\ref{b2}) of the term $\dfrac{\partial^2 \alpha_\nu}{\partial {\dot q}_r \partial t}$ for (NNC) systems or of the term $\dfrac{\partial \alpha_{\nu,r}}{\partial t}$ for (LNC).
		
\subsubsection*{ $\check{\rm C}$aplygin's systems} 
A special case which is recurrent in literature and applications concerns stationary (LNC) 
systems with the additional assumptions
\begin{equation}
	\begin{array}{ll}
		\label{chaplin}
		\alpha_{\nu,j}=\alpha_{\nu,j} (q_1, \dots, q_m)& \textrm{for each}\; \nu=1,\dots, k\;\textrm{and}\;j=1, \dots,m\\
		\\\
		T=T(q_1, \dots, q_m, {\dot q}_1, \dots, {\dot q}_n), & \mathcal{F}_i=\mathcal{F}_i(q_1, \dots, q_m, {\dot q}_1, \dots, {\dot q}_n)
	\end{array}
\end{equation}
Equations (\ref{vnl}) reduce to 
\begin{equation}
	\label{capligineq}
	\dfrac{d}{dt}\dfrac{\partial T^*}{\partial {\dot q}_r}-\dfrac{\partial T^*}{\partial q_r}
	-\sum\limits_{\nu=1}^k\sum\limits_{j=1}^m 
	\left(\dfrac{\partial \alpha_{\nu, r}}{\partial q_j}-
	\dfrac{\partial \alpha_{\nu,j}}{\partial q_r}\right){\dot q}_j
	\dfrac{\partial T}{\partial {\dot q}_{m+\nu}}=
	{\cal F}^{(q_r)}+\sum\limits_{j=1}^{k} \alpha_{j,r} {\cal F}^{(q_{m+j})}
	\quad r=1,\dots, m.
\end{equation}	
called $\check{\rm C}$aplygin's equations (dating (\cite{capligin}), see also \cite{neimark}).
The evident advantage consists is that (\ref{capligineq}) contains only the unknown functions $q_1$, $\dots$, $q_m$ and it is disentangled from the constraints equations (\ref{constrlinexpl}).

\subsubsection*{Further special forms}
Let us underline the following aspects.
\begin{itemize}
	\item[$(a)$]

If $T$ is of the form $T=T({\dot q}_1, \dots, {\dot q}_n)$, then the left part of (\ref{vnl}) reduces to 
$\dfrac{d}{dt}\dfrac{\partial T^*}{\partial {\dot q}_r}-\dfrac{\partial T^*}{\partial q_r}
-\sum\limits_{\nu=1}^k  B_r^\nu \dfrac{\partial T}{\partial {\dot q}_{m+\nu}}$. 

\item[$(b)$]
On the other hand, if the conditions (\ref{constr}) for a (NNC) system are $\phi_\nu({\dot q}_1, \dots, {\dot q}_n,t)=0$, $\nu=1, \dots, k$, then (\ref{constrexpl}) are of the form
\begin{equation}
	\label{noq}
	\alpha_\nu=\alpha_\nu({\dot q}_1, \dots, {\dot q}_m, t), \;\;\;\nu=1, \dots, k.
\end{equation}
In this case, the coefficients (\ref{b}) are simply 
$B_r^\nu=\dfrac{d}{dt}
\left(
\dfrac{\partial \alpha_\nu}{\partial {\dot q}_r}\right)$.
\end{itemize}

\noindent
Putting together $(a)$ and $(b)$, we deduce that if both the conditions on $T$ and on $\phi_\nu$ are verified, then the equations of motion (\ref{vnl}) are simplified as

\begin{equation}
\label{vnlab}
\dfrac{d}{dt}\dfrac{\partial T^*}{\partial {\dot q}_r}-
\sum\limits_{\nu=1}^k  
\dfrac{d}{dt}
\left(
\dfrac{\partial \alpha_\nu}{\partial {\dot q}_r}\right)
\dfrac{\partial T}{\partial {\dot q}_{m+\nu}}
={\cal F}^{(q_r)}+\sum\limits_{\nu=1}^k \dfrac{\partial \alpha_\nu}{\partial {\dot q}_r} {\cal F}^{(q_{m+\nu})}, 
\qquad r=1,\dots, m.
\end{equation}

\noindent
We also remark that the equality 
\begin{equation}
	\label{tequiv}
	\dfrac{d}{dt}\dfrac{\partial T^*}{\partial {\dot q}_r}
	-\sum\limits_{\nu=1}^k  \dfrac{d}{dt}
	\left(
	\dfrac{\partial \alpha_\nu}{\partial {\dot q}_r}\right)
	\dfrac{\partial T}{\partial {\dot q}_{m+\nu}}=
	\dfrac{d}{dt}\dfrac{\partial T}{\partial {\dot q}_r}+
	\sum\limits_{\nu=1}^k
	\dfrac{\partial \alpha_\nu}{\partial {\dot q_r}}
	\dfrac{d}{dt}\left(
	\dfrac{\partial T}{\partial {\dot q}_{m+\nu}}\right)
\end{equation}
where the variables ${\dot q}_{k+1}, \dots, {\dot q}_n$ in $T$ have to be expressed according to (\ref{constrexpl}) after differentiation, 
is valid for any $r=1, \dots, m$ by virtue of (\ref{relder}). Hence, when (\ref{vnlab}) are applied, one of the two expressions in (\ref{tequiv}) can be considered, on the basis of convenience.

\noindent
In particular, if $T$ is the quadratic function 
\begin{equation}
	\label{tquadr}
T=\dfrac{1}{2}\sum\limits_{i=1}^n M^{(i)}{\dot q}_i^2,\qquad M^{(i)}>0
\end{equation}
(for instance in the case of cartesian coordinates) where $M^{(i)}$ is the mass pertaining to the $i$--th coordinate, then 
(\ref{trid}) takes the form
\begin{equation}
	\label{tstarcart}
	T^*=\dfrac{1}{2}\sum\limits_{i=1}^m M^{(i)}{\dot q}_i^2+\dfrac{1}{2}\sum\limits_{\nu=1}^k M^{(\nu)}\alpha_\nu^2
\end{equation}
hence $\dfrac{\partial T}{\partial {\dot q}_{m+\nu}}=M^{(m+\nu)}{\dot q}_{m+\nu}=M^{(m+\nu)}\alpha_\nu$
and (\ref{vnlab}) are, owing to (\ref{tequiv}),
\begin{equation}
\label{vnlquadr}
M^{(r)}{\q2dot\limits^{..}}_r+\sum\limits_{\nu=1}^k M^{(m+\nu)}\dfrac{\partial \alpha_\nu}
{\partial {\dot q_r}} \dfrac{d\alpha_\nu}{dt}=
{\cal F}^{(q_r)}+\sum\limits_{\nu=1}^{k} \dfrac{\partial \alpha_\nu}{\partial {\dot q}_r}{\cal F}^{(q_{m+\nu})}\quad r=1,\dots, m.
\end{equation}

\begin{exe}
The equations of motion for the point $P$ of mass $M$ subject to the constraint (\ref{pct}) and to the weight force are immediately written by means of (\ref{vnlquadr}):
$$
\begin{array}{l}
(C^2-{\dot q}_2^2){\q2dot^{..}}_1+{\dot q}_1
({\dot q}_2{\q2dot^{..}}_2-C{\dot C})=\pm g{\dot q}_1R, \\
\\
(C^2-{\dot q}_1^2){\q2dot^{..}}_2+{\dot q}_2
({\dot q}_1{\q2dot^{..}}_1-C{\dot C})=\pm g{\dot q}_2R
\end{array}
$$
where $R({\dot q}_1, {\dot q}_2,t)=C^2(t)-{\dot q_1^2-{\dot q}_2^2}$ (the equations are multiplied by $R^2/M$ and the derivative $\frac{d}{dt}$ is calculated explicitly).
\end{exe}

\begin{exe}
The constraints (\ref{quadrom}) with $a_{i,j}^{(\nu)}$ constant are of the type (\ref{noq})
and the Examples 2 (same magnitude of velocities), 3 (parallel velocities) and 4 (perpendicular velocities) show the kinetic energy of statement $(a)$:
concerning Example 2 and referring to the formulation (\ref{velparallexpl}), the equations of motion (\ref{vnlquadr}) are

\begin{equation}
	\label{eqsamevel}
\left\{
\begin{array}{ll}
M_1 {\q2dot^{..}}_r+{\dot q}_r\sum\limits_{j=2}^N \dfrac{M_j}{R_j}\dfrac{dR_j}{dt}=
{\cal F}^{(q_r)} \pm {\dot q}_r\sum\limits_{j=2}^N \dfrac{1}{R_j}{\cal F}^{(q_{2N+j})}, & r=1,2,3\\
\\
M^{(r)} {\q2dot^{..}}_r- \dfrac{{\dot q}_r}{R^{(r)}}\dfrac{dR^{(r)}}{dt}=
{\cal F}^{(q_r)} \mp \dfrac{{\dot q}_r}{R^{(r)}}{\cal F}^{(q_{2N+r})}
 & r=4,5,\dots, 2N+1
\end{array}
\right.
\end{equation}
where we set, for $j=2, \dots, N$ and $r=4,5,\dots, 2N+1$: 
$$
\begin{array}{lll}
R_j=\sqrt{{\dot q}_1^2+{\dot q}_2^2+{\dot q}_3^2-{\dot q}^2_{2j}-{\dot q}^2_{2j+1}} &
M^{(r)}=\left\{ \begin{array}{ll}
M_{r/2} &for\;r\;even\\
M_{(r-1)/2} & for\;r\;odd	
\end{array} \right.
&
R^{(r)}=\left\{ 
\begin{array}{ll}
R_{r/2} &for\;r\;even\\
R_{(r-1)/2} & for\;r\;odd	
\end{array} \right.
\end{array}
$$
Likewise, the case of $N$ point with parallel velocities of Example 3 is suitable for the use of the form (\ref{vnlquadr}): the jacobian matrix 
$\dfrac{\partial \alpha_\nu}{\partial {\dot q}_j}$, $\nu=1,\dots,2(N-1)$, $j=1,\dots, N+2$, according to the selection (\ref{velparallind}) of independent velocities (we recall $({\dot q}_1, \dots, {\dot q}_{N+2})=({\dot x}_1, \dots, {\dot x}_N, {\dot y}_1, {\dot z}_1)$) and to the explicit form (\ref{velparallexpl}), is
$$
J_{({\dot q}_1, \dots, {\dot q}_{N+2})}=\dfrac{1}{{\dot q}_1}
\left(
\begin{array}{cccc}
	-\dfrac{{\dot q}_{N+1}}{{\dot q}_1}
	{\dot {\bf q}}_{2,N}
&{\dot q}_{N+1} {\mathbb I}_{N-1} & {\dot {\bf q}}_{2,N} & {\bf 0}_{N-1}\\
	\\
-\dfrac{{\dot q}_{N+2}}{{\dot q}_1}	{\dot {\bf q}}_{2,N}
&{\dot q}_{N+2} {\mathbb I}_{N-1} & {\bf 0}_{N-1} &{\dot {\bf q}}_{2,N}
\end{array}
\right)
$$
where ${\mathbb I}_{N-1}$ is the identity matrix of order $N-1$, ${\dot {\bf q}}_{2,N}$ the column vector $\left(
\begin{array}{l}
	{\dot q}_2\\
	\dots \\
	\dots\\
	{\dot q}_N
\end{array}
\right)\in {\Bbb R}^{N-1}$ and ${\bf 0}_{N-1}$ the null column vector in ${\Bbb R}^{N-1}$.
The corresponding $N+2$ equations of motion (\ref{vnlquadr}) are immediately available:
\begin{equation}
	\label{eqparall}
\left\{
\begin{array}{l}
	M_1 {\q2dot^{..}}_1-\dfrac{1}{{\dot q}_1^2}\sum\limits_{j=2}^N M_j {\dot q}_j\left(
	{\dot q}_{N+1}
	\dfrac{d}{dt}\left(\dfrac{{\dot q}_{N+1}}{{\dot q}_1}{\dot q}_j\right)+
	{\dot q}_{N+2}
	\dfrac{d}{dt}\left(\dfrac{{\dot q}_{N+2}}{{\dot q}_1}{\dot q}_j\right)
	\right)=\\
	\\
	\quad
	{\cal F}^{(q_1)} -\dfrac{1}{{\dot q}_1^2}
\sum\limits_{j=2}^N {\dot q}_j \left({\dot q}_{N+1}{\cal F}^{(q_{N+1+j})}+
	{\dot q}_{N+2}{\cal F}^{(q_{2N+j})}\right)
\\
\\
M_r {\q2dot^{..}}_r+M_r \left(
\dfrac{{\dot q}_{N+1}}{{\dot q}_1}\dfrac{d}{dt}\left(\dfrac{{\dot q}_{N+1}}{{\dot q}_1}{\dot q}_r\right)+
\dfrac{{\dot q}_{N+2}}{{\dot q}_1}\dfrac{d}{dt}\left(\dfrac{{\dot q}_{N+2}}{{\dot q}_1}{\dot q}_r\right)
\right)
=\\
\\
\quad {\cal F}^{(q_r)} +\dfrac{1}{{\dot q}_1}\left({\dot q}_{N+1}{\cal F}^{(q_{N+1+r})}+
	{\dot q}_{N+2}{\cal F}^{(q_{2N+r})}\right)
	\;
	\quad r=2, 3, \dots, N\\
	\\
M_1 {\q2dot^{..}}_{N+1}+\sum\limits_{j=2}^N  M_j\dfrac{{\dot q}_j}{{\dot q}_1}
\dfrac{d}{dt}\left(\dfrac{{\dot q}_j}{{\dot q}_1}{\dot q}_{N+1}\right)=
{\cal F}^{(q_{N+1})}+\dfrac{1}{{\dot q}_1}\sum\limits_{j=2}^N {\dot q}_j {\cal F}^{(q{N+1+j})}
	\\
	\\
M_1 {\q2dot^{..}}_{N+2} +\sum\limits_{j=2}^N M_j\dfrac{{\dot q}_j}{{\dot q}_1}
\dfrac{d}{dt}\left(\dfrac{{\dot q}_j}{{\dot q}_1}{\dot q}_{N+2}\right)=
{\cal F}^{(q_{N+2})}+\dfrac{1}{{\dot q}_1}\sum\limits_{j=2}^N {\dot q}_j {\cal F}^{(q_{2N+j})}
\end{array}
\right.
\end{equation}
\end{exe}

\begin{exe}
A similar scheme can be obtained in the case of Example 4 (perpendicular velocities), if one considers the contraints (\ref{velperp}): for $N=2$ (hence $k=1$ and $m=5$), setting $(x_1, y_1, z_1, x_2, y_2, z_2)=(q_1, q_2, q_3, q_4, q_5, q_6)$ the only constraint can be put in the form (\ref{constrexpl}) as ${\dot q}_6=\alpha_1({\dot q}_1, {\dot q}_2, {\dot q}_3, {\dot q}_4, {\dot q}_5)=-\dfrac{{\dot q}_1 {\dot q}_4+{\dot q}_2 {\dot q}_5}{{\dot q}_3}$ and equations (\ref{vnlquadr}) are
$$
\left\{
\begin{array}{l}
M_1{\q2dot^{..}}_1-M_2\dfrac{{\dot q}_4}{{\dot q}_3}\dfrac{d\alpha_1}{dt} =
 {\cal F}^{(q_1)} -\dfrac{{\dot q}_4}{{\dot q}_3} {\cal F}^{(q_6)}\\
\\
M_1{\q2dot^{..}}_2-M_2\dfrac{{\dot q}_5}{{\dot q}_3}\dfrac{d\alpha_1}{dt}=  {\cal F}^{(q_1)} 
-\dfrac{{\dot q}_5}{{\dot q}_3}   {\cal F}^{(q_6)}\\
\\
M_1{\q2dot^{..}}_3+M_2\dfrac{\alpha_1}{{\dot q}_3}\dfrac{d\alpha_1}{dt} =  {\cal F}^{(q_1)} -\dfrac{\alpha_1}{{\dot q}_3}    {\cal F}^{(q_6)}\\
\\
M_2{\q2dot^{..}}_4-M_2\dfrac{{\dot q}_1}{{\dot q}_3}\dfrac{d\alpha_1}{dt} =  {\cal F}^{(q_1)} -\dfrac{{\dot q}_1}{{\dot q}_3}    {\cal F}^{(q_6)}\\
\\
M_2{\q2dot^{..}}_5-M_2\dfrac{{\dot q}_2}{{\dot q}_3}\dfrac{d\alpha_1}{dt} =  {\cal F}^{(q_1)} -\dfrac{{\dot q}_2}{{\dot q}_3}    {\cal F}^{(_6)}
\end{array}
\right.
$$
where $M_1$ and $M_2$ are the masses of the two points.
\end{exe}

\subsubsection*{The equations of motion via the acceleration vector}
An alternative formal way to achieve the equations of motion (\ref{hlnc}) consists in calculating directly the acceleration of the system and the scalar product with ${\bf X}_r$: namely, recalling (\ref{vellagr}), (\ref{xm}) and taking into account (\ref{constrexpl}), one has

\begin{eqnarray*}
{\dot {\bf Q}}&=&
\sum\limits_{r=1}^m \left[
\left(
\dfrac{\partial {\bf X}^{(\mathsf{M})}}{\partial q_r}
+\sum\limits_{\nu=1}^k\dfrac{\partial {\bf X}^{(\mathsf{M})}}{\partial q_{m+\nu}}\dfrac{\partial \alpha_\nu}{\partial {\dot q}_r} 
\right) {\q2dot^{..}}_r \right.\\
\\
&+&
\left. \sum\limits_{s=1}^m \dfrac{\partial^2 {\bf X}^{(\mathsf{M})}}{\partial q_r \partial q_s}
{\dot q}_r {\dot q}_s 
+
\sum\limits_{\nu=1}^{k} \left(
2\dfrac{\partial^2 {\bf X}^{(\mathsf{M})}}{\partial q_r \partial q_{m+\nu}}\alpha_\nu+
\dfrac{\partial {\bf X}^{(\mathsf{M})}}{\partial q_{m+\nu}}
\dfrac{\partial \alpha_\nu}{\partial q_r} \right)+
2\dfrac{\partial^2 {\bf X}^{(\mathsf{M})}}{\partial q_r \partial t}
\right] {\dot q}_r\\
\\
&+&
\sum\limits_{\nu,\mu=1}^k \left(
\dfrac{\partial^2 {\bf X}^{(\mathsf{M})}}{\partial q_{m+\nu} \partial q_{m+\mu}}\alpha_\nu \alpha_\mu
+
\dfrac{\partial {\bf X}^{(\mathsf{M})}}{\partial q_{m+\nu}}
\dfrac{\partial \alpha_\nu}{\partial q_{m+\mu}}\alpha_\mu
\right)
+
\sum\limits_{\nu=1}^k\left(2 \dfrac{\partial^2 {\bf X}^ {(\mathsf{M})}}{\partial q_{m+\nu} \partial t}\alpha_\nu+ 
\dfrac{\partial {\bf X}^{(\mathsf{M})}}{\partial q_{m+\nu}}\dfrac{\partial \alpha_\nu}{\partial t} \right)+
\dfrac{\partial^2 {\bf X}^{(\mathsf{M})}}{\partial t^2}
\end{eqnarray*}
and $({\dot {\bf Q}}-{\bf F})\cdot {\bf X}_r=0$ for (NNC) systems corresponds to (details can be found in
\cite{tal})

\begin{equation}
	\label{vnl2}
	\sum\limits_{j=1}^m \left( C_r^j {\q2dot^{..}}_j+
	\sum\limits_{k=1}^m D_r^{j,k} {\dot q}_j{\dot q}_k + E_r^j {\dot q}_j\right) 
	+G_r={\cal F}^{(q_r)}+
	\sum\limits_{\nu=1}^k \dfrac {\partial \alpha_\nu}{\partial {\dot q}_r} {\cal F}^{(q_{m+\nu})}, 
	\qquad r=1,\dots, m
\end{equation}
where the coefficients 
$C_r^j$, $D_r^{j,k}$, $E_r^j$ and $G_r$, $r,j,k=1,\dots, m$, depending on $q_1, \dots, q_n$, ${\dot q}_1$, $\dots$, ${\dot q}_m$, $t$, are defined by

\begin{eqnarray}
	\nonumber
	C_r^j&=&g_{r,j}+\sum\limits_{\nu,\mu=1}^k\left(g_{r,m+\nu}
	\dfrac{\partial \alpha_\nu}{\partial {\dot q}_j}
	+g_{m+\nu,j}\dfrac{\partial \alpha_\nu}{\partial {\dot q}_r}+ g_{m+\nu,m+\mu}\dfrac{\partial \alpha_\mu}{\partial {\dot q}_r}\dfrac{\partial \alpha_\nu}{\partial {\dot q}_j}\right) \\
\label{coeffcdef}
	D_r^{j,k}&=&\xi_{j,k,r}+\sum\limits_{\nu=1}^k\xi_{j,k,m+\nu}\dfrac{\partial \alpha_\nu}{\partial {\dot q}_r} \\
		\nonumber	
	E_r^j &=&  \sum\limits_{\nu,\mu=1}^k \left(
	2\xi_{j,m+\nu,m+\mu} \alpha_\nu + g_{m+\nu,m+\mu}\dfrac{\partial \alpha_\nu}{\partial q_j}
	+2 \eta_{j,m+\mu} \right)\dfrac{\partial \alpha_\mu}{\partial {\dot q}_r}
	+\sum\limits_{\nu=1}^k \left( 2\xi_{j,m+\nu,r} \alpha_\nu + 
	g_{m+\nu, r}\dfrac{\partial \alpha_\nu}{\partial q_j} \right)+2 \eta_{j,r} \\
	\nonumber
	G_r&=&
	\sum\limits_{\nu,\mu,p=1}^k
	\left( \alpha_\mu\left( \xi_{m+\nu, m+\mu,m+p}\alpha_\nu + g_{m+\nu,m+p}
	\dfrac{\partial \alpha_\nu}{\partial q_{m+\mu}}\right)+
	2\eta_{m+\nu, m+p}\alpha_\nu +g_{m+\nu, m+p} \dfrac{\partial \alpha_\nu}{\partial t}+\zeta_{m+p}\right) \dfrac{\partial \alpha_p}{\partial {\dot q}_r} \\
	\nonumber
	&+&
	\sum\limits_{\nu,\mu=1}^k\alpha_\mu \left(
	\xi_{m+\nu,m+\mu,r}\alpha_\nu+g_{m+\nu,r} \dfrac{\partial \alpha_\nu}{\partial q_{m+\mu}}
	\right) +\sum\limits_{\nu=1}^k \left( 
	2 \eta_{m+\nu,r} \alpha_\nu +g_{m+\nu,r} \dfrac{\partial \alpha_\nu}{\partial t}\right)+\zeta_r.
\end{eqnarray}
with $g_{r,j}$ the same as in (\ref{gbc}) and, for any $a,b,c=1, \dots, n$:

\begin{equation}
	\label{gxi}
	\begin{array}{lll}
	\xi_{a,b,c}=\dfrac{\partial^2{\bf X}^{(\mathsf{M})}}{\partial q_a\partial q_b}\cdot \dfrac{\partial {\bf X}}{\partial q_c}, & 
		\eta_{a,b}= \dfrac{\partial^2 {\bf X}^{(\mathsf{M})}}{\partial q_a \partial t}\cdot \dfrac{\partial {\bf X}}{\partial q_b}, &
		\zeta_a=\dfrac{\partial^2 {\bf X}^{(\mathsf{M})}}{\partial t^2}\cdot 
		\dfrac{\partial {\bf X}}{\partial q_a}.
	\end{array} 
\end{equation}	

\noindent
It is worth noting that the expression that multiplies ${\q2dot^{..}}_r$ in the calculation of ${\dot {\bf Q}}$ is exactly the vector ${\bf X}_r$ (see (\ref{hlnc})) multiplied by the masses of the points: 
on the other hand this is the only term of the acceleration vector in which the second derivatives ${\q2dot^{..}}_r$ appear: this allows us to write, if we define
$S=\frac{1}{2} {\dot {\cal Q}}\cdot \xgrande2dot^{..}$, where ${\bf X}$ is the representative vector (\ref{xqt}), as the acceleration energy or Gibbs--Appell function:
$$
\dfrac{\partial S}{\partial {\q2dot^{..}}_r}=
{\dot {\cal Q}}\cdot \dfrac{\partial \xgrande2dot^{..}}{\partial {\q2dot^{..}}_r}={\dot {\cal Q}}\cdot
\left(\frac{\partial {\bf X}}{\partial q_r}+
\sum\limits_{\nu=1}^{k}\frac{\partial \alpha_\nu}{\partial {\dot q}_r}\frac{\partial {\bf X}}{\partial q_{m+\nu}}\right).
$$
This means that we can formally write the equations of motion (\ref{vnl2}) for (NNC) systems in the equivalent form 
$$
({\dot {\bf Q}}-{\bf F})\cdot {\bf X}_r=0\qquad \Leftrightarrow \qquad \dfrac{\partial S}{\partial {\q2dot^{..}}_r}={\cal F}^{(q_r)}+
\sum\limits_{\nu=1}^k \dfrac {\partial \alpha_\nu}{\partial {\dot q}_r} {\cal F}^{(q_{m+\nu})}, 
\qquad r=1,\dots, m
$$
The latter write is known as Gibbs--Appell equations. This extremely compact and general form to be linked to the gauss principle appears in \cite{appell1} and is developed
in \cite{gant} and \cite{papastrav}.

\noindent
In the applications, the explicit form (\ref{vnl2}) is frequently more accessible compared to (\ref{vnl}): actually, once the coefficients (\ref{gxi}) are known, only the derivatives of the functions $\alpha_\nu$ have to be calculated.
Moreover, equations (\ref{vnl}) contain many redundant (in the sense of deleting each other)
terms: it can be checked (see \cite{tal})) that all the terms of $-\sum\limits_{\nu=1}^k \frac{\partial T}{\partial {\dot q}_{m+\nu}} B_r^\nu$, $=1,\dots, m$, cancel out with part of the addends of $\frac{d}{dt}\frac{\partial T^*}{\partial {\dot q}_r}$,  
of $-\frac{\partial T^*}{\partial q_r}$ and of
$-\sum\limits_{\nu=1}^k\frac{\partial T^*}{\partial q_{m+\nu}}\frac{\partial \alpha_\nu}{\partial {\dot q_r}}$. 
We also notice that in case of scleronomic holonomic constraints ${\bf X}({\bf q})$ (see (\ref{xqt})) we get $\eta_{a,b}=0$, $\zeta_a=0$ for any $a,b=1, \dots, n$. Furthermore, in the absence of geometric constraints and using the $3N$ cartesian coordinates for $(q_1, \dots, q_n)$, it is $g_{r,r}=M^{(r)}$, with $M^{(r)}$ the mass of the point which the coordinate $q_r$ refers to, $g_{r,j}=0$ for $r\not =j$ and all the quantities in (\ref{gxi}) are null. Hence the coefficients (\ref{coeffcdef}) are
\begin{equation}
\label{coeffrid}
\begin{array}{l}
C_r^r= M^{(r)}+\sum\limits_{\nu=1}^k M^{(m+\nu)}
\left(\dfrac{\partial \alpha_\nu}{\partial {\dot q}_r}\right)^2, \quad 
C_r^j= \sum\limits_{\nu=1}^k M^{(m+\nu)}
\dfrac{\partial \alpha_\nu}{\partial {\dot q}_r}\dfrac{\partial \alpha_\nu}{\partial {\dot q}_j}\;\;
r\not=j \\
D_r^{j,k}=0, \quad 
E_r^j=\sum\limits_{\nu=1}^k M^{(m+\nu)}\dfrac{\partial \alpha_\nu}{\partial q_j}
\dfrac{\partial \alpha_\nu}{\partial {\dot q}_r}, 
\quad
G_r=
\sum\limits_{\nu,\mu=1}^k M^{(m+\nu)}
\alpha_\mu\dfrac{\partial \alpha_\nu}{\partial q_{m+\mu}}
\dfrac{\partial \alpha_\nu}{\partial {\dot q}_r}.   
\end{array}
\end{equation}

\begin{exe}
The case of the nonholonomic pendulum (Example 5) fits for th just described procedure: rewriting (\ref{nhpexpl}) as 
 	
 $$
\alpha_1(q_1, q_2, q_3, q_4, {\dot q}_1, {\dot q}_2, {\dot q}_3)=\dfrac{b(q_1{\dot q}_1+q_2{\dot q_2})-
	q_3(bq_1-aq_2)}{a(q_1{\dot q}_1+q_2{\dot q_2})+q_4(bq_1-aq_2)}{\dot q}_3=\dfrac{N(q_1, q_2, q_3,{\dot q}_1, {\dot q}_2)}{D(q_1, q_2,q_4, {\dot q}_1, {\dot q}_2)}{\dot q}_3
 $$
 and defining $P(q_1, q_2, q_3, q_4)= (aq_1+bq_2)(aq_3+bq_4)$, one has 
 $$
\dfrac{\partial \alpha_1}{\partial {\dot q}_1}=-{\dot q}_2{\dot q}_3\dfrac{P}{D^2}, \quad
\dfrac{\partial \alpha_1}{\partial {\dot q}_2}={\dot q}_1{\dot q}_3\dfrac{P}{D^2}, \quad
\dfrac{\partial \alpha_1}{\partial {\dot q}_3}=\dfrac{N}{D}
$$
and the calculation of (\ref{coeffrid}) allows to write the equations of motion (\ref{vnl2}) 
in the form
$$
\begin{array}{l}
M^{(1)}{\q2dot^{..}}_1+{\dot q}_2{\dot q}_3\left(\Psi+\dfrac{P}{D^2}{\cal F}^{(q_4)}\right)={\cal F}^{(q_1)}
\\
\\
M^{(2)}{\q2dot^{..}}_2-{\dot q}_1{\dot q}_3\left(\Psi+\dfrac{P}{D^2} {\cal F}^{(q_4)}\right)={\cal F}^{(q_2)}

\\
\\
M^{(3)}{\q2dot^{..}}_3-\dfrac{ND}{P}\left(\Psi+\dfrac{P}{D^2} {\cal F}^{(q_4)}\right)
={\cal F}^{(q_3)}
\end{array}
$$
 	
$$
\begin{array}{l}
 \Psi(q_1, q_2, q_3, q_4, {\dot q}_1, {\dot q}_2, {\dot q}_3, {\q2dot^{..}}_1, {\q2dot^{..}}_2, {\q2dot^{..}}_3)=\\
 \\
M^{(4)}\dfrac{P}{D^4}
\left[P{\dot q}_3({\dot q}_2{\q2dot^{..}}_1
-{\dot q}_1{\q2dot^{..}}_2)-ND{\q2dot^{..}}_3
-(b{\dot q}_1-a{\dot q}_2)\left((aq_3+bq_4)({\dot q}_1^2+{\dot q}_2^2)-{\dot q}_3\left(D-\dfrac{N}{D}{\dot q}_3\right)\right)
\right]
\end{array}
$$ 	
The equations of motion are simply and promptly obtained by means of (\ref{coeffrid}) in comparison with other methods and they appear arranged in the correct way in order to search for solutions of specific type.
\end{exe}

\begin{exe} 
	A particular example, analysed in \cite{virga}, outlines the fact that equations (\ref{vnl}) are still valid even though the function $T^*$ defined in (\ref{tmobile}) degenerates w.~r.~t.~the restricted variables ${\dot q}_1$, $\dots$, ${\dot q}_m$: let $P$ be a point of mass $M$ and cartesian coordinates $(x,y,z)$ and whose velocity is constant in module: 
	$|{\dot P}|=C>0$. 
In terms of $(q_1, q_2, q_3)=(x,y,z)$ we write (\ref{constrexpl})
	as ${\dot q}_3=\pm \sqrt{C^2-{\dot q}_1^2 - {\dot q}_2^2}=\alpha_1({\dot q}_1, {\dot q}_2)$, where the sign depends on the initial conditions. 
	Concerning (\ref{vnl2}), we use (\ref{coeffrid}) with $g_{1,1}=g_{2,2}=g_{3,3}=m$, $g_{i,j}=0$ for $i\not =j$ and the only nonzero coefficients (since $\alpha_1$ does not depend on $q_1$, $q_2$, $q_3$) are
	$$
	C_i^i= M\left(
	1+\frac{M{\dot q}_i^2}{C^2-{\dot q}_1^2-{\dot q}_2^2}\right), \quad i=1,2, \qquad  
	C_2^1=C_1^2=
	\frac{M {\dot q}_1 {\dot q}_2}{C^2- {\dot q}_1^2 - {\dot q}_2^2}.
	$$
	Hence (\ref{vnl2}) are
	\begin{equation}
		\label{velcost}
		\left\{
		\begin{array}{l}
			M \frac{C^2-{\dot q}_2^2}{C^2-{\dot q}_1^2 -{\dot q_2^2}}{\q2dot^{..}}_1
			+ M \frac{{\dot q}_1 {\dot q}_2}{C^2-{\dot q}_1^2 -{\dot q_2^2}}{\q2dot^{..}}_2
			= {\cal F}^{(q_1)}\mp {\cal F}^{(q_3)} 
			\frac{{\dot q}_1}{\sqrt{C^2-{\dot q}_1^2 - {\dot q}_2^2}}, \\ 
			M \frac{{\dot q}_1 {\dot q}_2}{C^2-{\dot q}_1^2 -{\dot q}_2^2}{\q2dot^{..}}_1+
			M \frac{C^2-{\dot q}_1^2}{C^2-{\dot q}_1^2 -{\dot q}_2^2}{\q2dot^{..}}_2
			= {\cal F}^{(q_1)}\mp {\cal F}^{(q_3)} 
			\frac{{\dot q}_2}{\sqrt{C^2-{\dot q}_1^2 - {\dot q}_2^2}},	
		\end{array}
		\right.
	\end{equation}
	On the other hand, if the equations are written by means of (\ref{vnl}), one has that (\ref{tmobile}) is $T^*=\frac{1}{2}mC^2$, hence the only contributions to the left--hand side terms of (\ref{vnl}) are 
	$$
	-B^1_i\dfrac{\partial T}{\partial {\dot q}_3}=
	-m\left(\dfrac{\partial^2 \alpha_1}{\partial {\dot q}_i^2}{\q2dot^{..}}_1 
	+\dfrac{\partial^2 \alpha_1}{\partial {\dot q}_i \partial {\dot q}_2}{\q2dot^{..}}_2\right){\dot q}_3({\dot q}_1, {\dot q}_2), \quad i=1, 2
	$$
	Once the second derivatives are calculated, equations (\ref{velcost}) are found again. We remark that the more general case $C=C(t)$ has been treated in Example 11: in the present example the attention is drawn to the comparison between (\ref{vnl}) and (\ref{vnl2}).
\end{exe}

\subsubsection*{On the selection of independent velocities}

\noindent
Let us finally investigate the role 
of a certain selection of the independent velocities $({\dot q}_1, \dots, {\dot q}_m)$, compared to a second choice of $m$--uple. We consider worthwhile to examine such a focused question, 
rather than introducing a general change of the lagrangian coordinates $(q_1, \dots, q_n)$. 
Hence, assuming that a second explicit set is deducible from (\ref{constr}):

\begin{equation}
	\label{constrexpl2}
	\begin{cases}
		{\dot q}_{\sigma_{m+1}}=\alpha_1^{(\sigma)}(q_1, \dots, q_n, {\dot q}_{\sigma_1}, \dots, {\dot q}_{\sigma_m}, t) \\
		\dots  \\
		{\dot q}_{{\sigma}_{m+k}}=\alpha_k^{(\sigma)}(q_1, \dots, q_n, {\dot q}_{\sigma_1}, \dots, {\dot q}_{\sigma_m}, t)
	\end{cases}
\end{equation}
where $({\dot q}_{\sigma_1}, \dots, {\dot q}_{\sigma_m}, {\dot q}_{\sigma_{m+1}}, \dots, {\dot q}_{\sigma_n})$ is a permutation of $({\dot q}_1, \dots, {\dot q}_n)$ fulfilling 
$det\,J_{({\dot q}_{\sigma_{m+1}}, \dots, {\dot q}_{\sigma_n})}(\Phi_1, \dots, \Phi_k)\not =0$.

\noindent
The independent velocities $({\dot q}_1, \dots, {\dot q}_m)$ and the dependent kinetic variables (\ref{constrexpl}) ${\dot q}_{m+1}=\alpha_1$, $\dots$, ${\dot q}_n=\alpha_k$ can be splitted on the basis of:
\begin{description}
	\item[] $({\dot q}_1, \dots, {\dot q}_\ell)$, $0\leq \ell \leq m$, are the velocities among $({\dot q}_1, \dots, {\dot q}_m)$ which remain independent parameters, say  $({\dot q}_{\sigma_1}, \dots, {\dot q}_{\sigma_\ell})$ in the same order, without loss of generality,
	
	\item[]	$({\dot q}_{\ell+1}, \dots, {\dot q}_m)$ turn into the $m-\ell$ dependent variables $({\dot q}_{\sigma_{m+h+1}}, \dots, {\dot q}_{\sigma_{m+k}})$, where $k-h=m-\ell$, 
	
	\item[] $({\dot q}_{m+1}, \dots, {\dot q}_{m+h})$ remain the dependent variables $({\dot q}_{\sigma_{m+1}}, \dots, {\dot q}_{\sigma_{m+h}})$, 
	
	\item[] $({\dot q}_{m+h+1}, \dots, {\dot q}_{m+k})$ turn into the $k-h=m-\ell$ independent velocities $( {\dot q}_{\sigma_{\ell+1}}, \dots, {\dot q}_{\sigma_{m}})$.
	
\end{description}

\noindent
Consequently, the position ${\dot q}_i={\dot q}_i({\dot q}_{\sigma_1}, \dots, {\dot q}_{\sigma_m})$, $i=1, \dots, m$ is
\begin{equation}
	\label{unoelle}
	\begin{array}{l}
		{\dot q}_1={\dot q}_{\sigma_1},\;\; \dots\;\; {\dot q}_\ell={\dot q}_{\sigma_\ell}, \\
		{\dot q}_{\ell+1}=\alpha^{(\sigma)}_{\sigma_{h+1}}({\dot q}_{\sigma_1}, \dots, {\dot q}_{\sigma_m}),\;\; \dots \;\;{\dot q}_m=\alpha^{(\sigma)}_{\sigma_k}({\dot q}_{\sigma_1}, \dots, {\dot q}_{\sigma_m}).
	\end{array}
\end{equation}
At the same time:
\begin{equation}
	\label{ellepiuunom}
	\begin{array}{l}
		\alpha_1=\alpha_1^{(\sigma)}\;\; \dots\;\;  \alpha_h= \alpha_h^{(\sigma)}, \\
		{\dot q}_{\sigma_{\ell+1}}=\alpha_{h+1}({\dot q}_1, \dots, {\dot q}_m), \;\;\dots \;\;
		{\dot q}_{\sigma_{m}}=\alpha_k ({\dot q}_1, \dots, {\dot q}_m).
	\end{array}
\end{equation}

\noindent
We refer to $\ell=0$ as the case when all of the original kinetic variables become dependent; the case $\ell=m$ is trivial.

\noindent
The relation between the two sets of equations can be expressed in terms of the transposed jacobian matrix

\begin{equation}
	\label{iacvel}
	\begin{array}{l}
		J^T_{({\dot q}_{\sigma_1}, \dots, {\dot q}_{\sigma_m})}({\dot q}_1, \dots {\dot q}_m)=
		\left(
		\dfrac{\partial {\dot q}_j}{\partial {\dot q}_{\sigma_i}}
		\right)_{i,j=1, \dots, m}
		=
		\left(
		\begin{array}{cccc}
			&  
			\frac{\partial \alpha_{h+1}^{(\sigma)}}{\partial {\dot q}_{\sigma_1}} & \dots & 
			\frac{\partial \alpha_k^{(\sigma)}}{\partial {\dot q}_{\sigma_1}}	
			\\
			{\Bbb I}_\ell	 &\dots  & &\dots \\
			\\
			{\Bbb O}_{(m-\ell)\times \ell} & 
			\frac{\partial \alpha_{h+1}^{(\sigma)}}{\partial {\dot q}_{\sigma_m}} & \dots & 
			\frac{\partial \alpha_k^{(\sigma)}}{\partial {\dot q}_{\sigma_m}}	
		\end{array}
		\right)
	\end{array}
\end{equation}
where ${\Bbb I}_{\ell}$ is the identity matrix of size $\ell$, ${\Bbb O}_{(m-\ell)\times \ell}$ is the $(m-\ell)\times \ell$--null matrix and the functions appearing in the entries are those of (\ref{constrexpl2}).
More precisely, it is not difficult to prove the following

\begin{prop}
	The equations of motion written with the selection $({\dot q}_{\sigma_1}, \dots, {\dot q}_{\sigma_m})$ as independent ve\-lo\-ci\-ties are obtained by left multiplying the $m$ vector of equations of motion relative to $({\dot q}_1, \dots, {\dot q}_m)$ by the matrix (\ref{iacvel}). That is, if $E_i=0$, $i=1, \dots, m$, are the $m$ equations (\ref{vnl}) (with the force terms moved to the left side), thw equations of motion $E_{\sigma_i}=0$ corresponding to the setting (\ref{constrexpl2}) verify
	\begin{equation}
		\label{eqsigma}
		E_{\sigma_i}=\sum\limits_{i=1}^m \dfrac{\partial {\dot q}_r}{\partial {\dot q}_{\sigma_i}}E_r, \quad i=1, \dots, m.
	\end{equation}
\end{prop}

\section{Conclusions}

\noindent
The analysis of nonlinear kinematic constraints is certainly less debated than in the linear case, despite some very spontaneous constraint restrictions are naturally nonlinear (parallelism, perpendicularity,...).

\noindent
The present work aims to pursue a dual purpose:
\begin{itemize}
	\item[$(i)$] to give rise to a simple approach that generalizes the ordinary situation of the Eulero--Lagrange equations in the holonomic case, understood as Newton's equations projected along the directions of the possible speeds,
	
	\item[$(ii)$] to provide a set of equations that can be used to formulate examples and applications, keeping real speeds as kinetic variables, and exhibit a series of examples and applications for which this approach is congenial.
\end{itemize}

\noindent
As regards the first point, a proposal has been made regarding the description in terms of vectors of the possible displacements, extending what is known in the standard cases.
In the nonlinear case it is reasonable to remain in real coordinates, since the use of pseudo-velocity lends itself more easily to the case of linear transformations of kinetic variables.
For point $(ii)$ it must be said that the mere fact of writing the equations of motion for nonholonomic linear and nonlinear systems is anything but trivial, the procedures are almost always very complex.The procedure of writing the equations of motion is frequently faced with specific techniques, rather than with a systematic approach.

\noindent
We have also set ourselves the goal of taking care of an aspect that is sometimes treated superficially in the literature: at least in some examples, the effective equivalence of a condition or of a group of conditions has been examined in order to formulate the same binding situation.
In other cases we have taken care of writing the equations for decidedly recurring problems (the pursuing problem in the space, the nonholonomic pendulum, ...).
We have tried as much as possible to trace typologies of Lagrangians or constraints for which the calculation of the equations can be particularly shortened.

\noindent
The approach chosen predisposes to at least two themes of deepening the problem, which will be the next subjects of study:
\begin{itemize}
	\item[$(a)$] to generalize the class of constraints, also admitting the presence of higher-order derivatives,

	\item[$(b)$] to carry out the energy balance that follows from the equations, to examine the possibility of the presence of the integral of the energy, on the basis of certain hypotheses that the constraints and the applied forces must satisfy.
\end{itemize}

\end{document}